\documentclass[fleqn,usenatbib]{mnras}

\usepackage{newtxtext,newtxmath}

\usepackage[T1]{fontenc}

\DeclareRobustCommand{\VAN}[3]{#2}
\let\VANthebibliography\thebibliography
\def\thebibliography{\DeclareRobustCommand{\VAN}[3]{##3}\VANthebibliography}

\usepackage{graphicx}	
\usepackage{amsmath}	
\usepackage{subcaption}

\newcommand{\mzero}{$m_{\mathrm 0}$}
\newcommand{\mone}{$m_{\mathrm 1}$}

\title[Diagnostics of lead stratification]{Spectroscopic diagnostics of lead stratification in hot subdwarf atmospheres}

\author[L. J. A. Scott et al.]{
L. J. A. Scott$^{1}$\thanks{E-mail: laura.scott@armagh.ac.uk}, C. S. Jeffery$^{1}$, C. M. Byrne$^{2}$ and M. Dorsch$^3$\\
\\
$^{1}$Armagh Observatory and Planetarium, College Hill, Armagh, BT61 9DB, UK\\
$^{2}$Department of Physics, University of Warwick, Gibbet Hill Road, Coventry CV4 7AL, UK\\
$^3$Dr. Karl Remeis-Observatory \& ECAP, Friedrich-Alexander University Erlangen-Nürnberg, Sternwartstr. 7, 96049 Bamberg,
Germany
}

\date{Accepted XXX. Received YYY; in original form ZZZ}

\pubyear{2023}

\begin{document}
\label{firstpage}
\pagerange{\pageref{firstpage}--\pageref{lastpage}}
\maketitle

\begin{abstract}
Heavy metal subdwarfs are a class of hot subdwarfs with very high abundances of heavy elements, typically around 10\,000 times solar. They include stars which are strongly enhanced in either lead or zirconium, as well as other elements. Vertical stratification of the enhanced elements, where the element is concentrated in a thin layer of the atmosphere, has been proposed as a mechanism to explain the apparent high abundances. This paper explores the effects of the vertical stratification of lead on theoretical spectra of hot subdwarfs. The concentration of lead in different regions of the model atmosphere is found to affect individual lines in a broadly wavelength-dependent manner, with the potential for lines to display modified profiles depending on the location of lead enhancement in the atmosphere. This wavelength dependence highlights the importance of observations in both the optical and the UV for determining whether stratification is present in real stars.
\end{abstract}




\section{Introduction}
The heavy metal subdwarfs are a rare class of stars characterised by extreme surface chemistries, with enhanced  helium and high abundances of heavy metals (e.g. zirconium and lead). They are a subset of hot subdwarfs, which are small stars ($\sim0.1$\,R$_{\odot}$, $\sim0.5$\,M$_{\odot}$) occupying a region of the Hertzsprung-Russell diagram at the blue end of the horizontal branch.

Whilst hot subdwarfs often have elevated metal abundances, the heavy metal subdwarfs are even more extreme \citep[see Fig.\,8 in][]{jeffery17a}. A prime example is EC\,22536--5304 \citep{dorsch21}, in which the atmospheric lead abundance is elevated by 8 dex compared to the metallicity of its cool dwarf companion. In addition, heavy metal subdwarfs can be split into a lead-rich group \citep[e.g.][]{dorsch19,naslim20,naslim13,wild18} and a zirconium-rich group \citep[e.g.][]{naslim11,latour19b,ostensen20}. The lead stars are higher in temperature, at around 37 to 42\,kK, compared to the zirconium stars at around 34 to 36\,kK. This raises the question of how the heavy metals came to be so concentrated in the stars' atmospheres, and what causes the temperature difference between the groups.

Chemical stratification can be disrupted by mixing processes such as convection and rotation. Hot subdwarf photospheres generally lack convection, except for hot helium-rich stars, in which a thin convective zone may form \citep{groth85}. The heavy metal subdwarfs are also slow rotators. The shortest rotational period for a heavy metal subdwarf has been measured at $\sim17$ days \citep{ostensen20}. In terms of projected rotational velocity, $v\sin{i}$, the fastest measured rotation is an upper limit of 10\,km\,s$^{-1}$ \citep{jeffery17a}.
Due to the lack of strong mixing processes, the atmospheres of heavy metal subdwarfs should be able to maintain chemical stratification.

One proposed mechanism to produce abundance stratification is the combination of radiative levitation and gravitational settling, which would concentrate elements into a region where the two processes are in equilibrium. These diffusive processes have been extensively studied in chemically peculiar stars, such as the Ap stars \citep{michaud70}, and have motivated the development of model atmospheres with stratified abundances \citep{leblanc09}. This also includes studies into horizontal abundance stratification across the star's surface, such as \citet{alecian21}. Other cases in which diffusive transport and chemical stratification have been considered include hot white dwarfs \citep[e.g. ][]{schuh02} and horizontal branch stars \citep{michaud08,michaud11}, as well as in the context of pulsation driving mechanisms in hot subdwarfs \citep{fontaine03,jeffery06c,byrne20}. Provided that the equilibrium between acceleration due to radiative levitation and gravitational settling occurs in the region of the atmosphere at which observable absorption lines are formed, this chemical transport would increase the measured abundances.

The questions of immediate concern are whether vertical chemical stratification can explain the abundances seen in heavy metal subdwarfs, and whether stratification can be identified by observable features in a star's spectrum.
In addition, how feasible is it to concentrate the necessary amount of lead into a thin layer without requiring additional nucleosynthesis?
These questions will be explored using model atmospheres with stratified lead abundance. Section\,\ref{sec:methods} will explain the models used, whilst Section\,\ref{sec:results} will describe the results. In Section\,\ref{sec:discussion}, the implications of the results will be discussed, with particular attention to observable fingerprints of stratification. Conclusions are given in Section\,\ref{sec:conclusions}.

\section{Methods}
\label{sec:methods}

\begin{figure*}
    \centering
    \includegraphics[width=\linewidth]{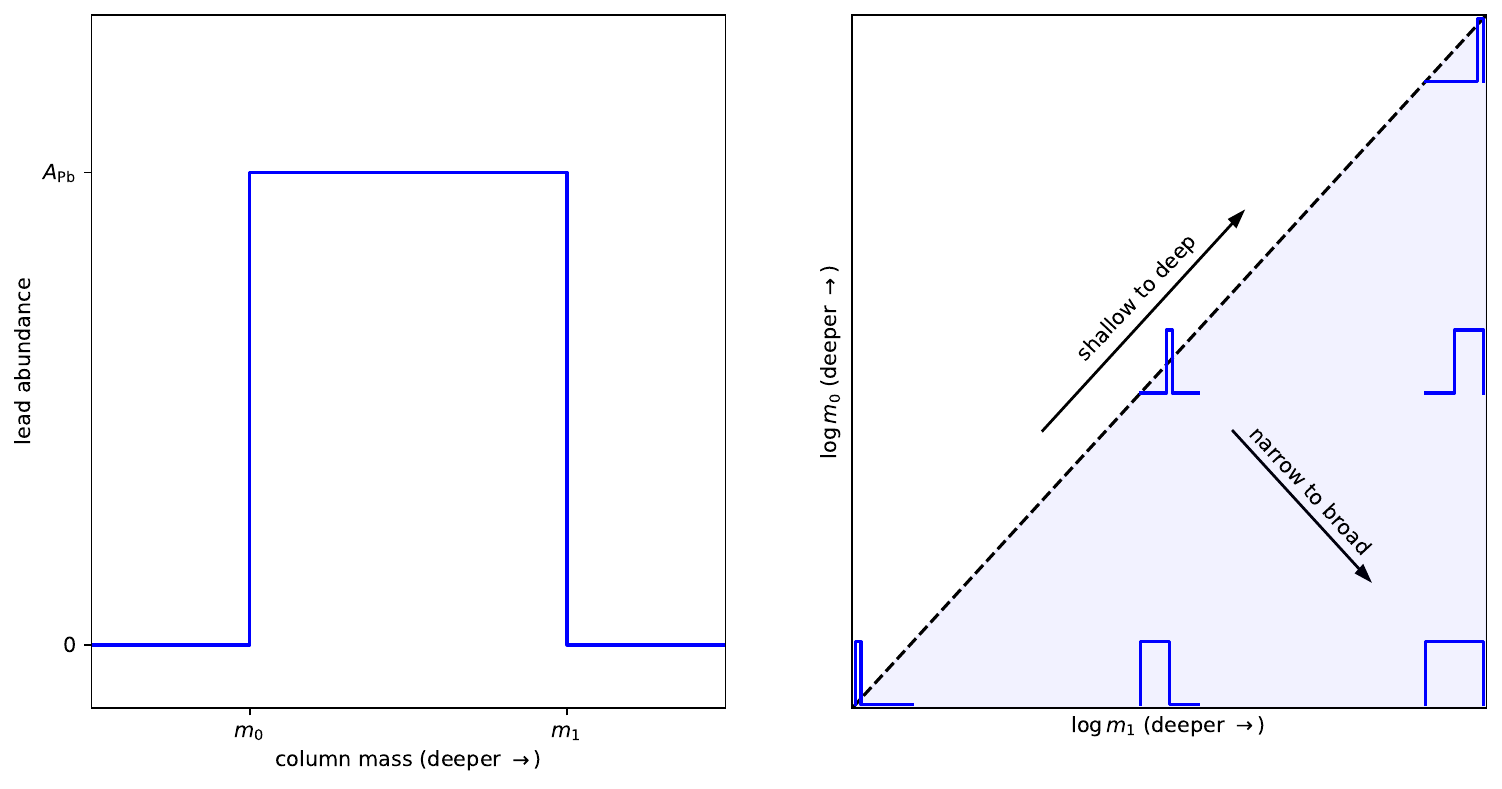}
    \caption{Schematic of the abundance profiles used in this study. The left-hand plot shows the shape of an individual profile, which consists of a lead-enriched layer located between a column mass of $m_{\rm 0}$ and $m_{\rm 1}$, where $m_{\rm 0} < m_{\rm 1}$. The lead abundance in the model atmosphere is $A_{\rm{Pb}}$ between $m_{\rm 0}$ and $m_{\rm 1}$, and 0 elsewhere. The right hand plot shows the full range of abundance profiles used. This is indicated by the shaded area below the dashed line, where $m_{\rm 0} < m_{\rm 1}$. At key points on the plane, the corresponding abundance profile is overplotted.}
    \label{fig:sketch}
\end{figure*}

\begin{table}
    \centering
    \caption{Parameters for the {\sc tlusty} models. $T_{\rm{eff}}$, $\log{g}$ and $n_{\rm{H}}$ refer to effective temperature, logarithm of surface gravity and number fraction of hydrogen respectively.}
    \label{tab:model_params}
    \begin{tabular}{l|c}
    \hline
        $T_{\rm{eff}}$ / kK & 34, 38, 42 \\
        $\log{g /{\mathrm{cm\,s^{-2}}}}$ & 5.80\\
        $n_{\rm{H}}$ & 0.5 \\
        Non-LTE ions & H\,{\sc i-ii}, He\,{\sc i-iii}\\
          & C\,{\sc i-v}, N\,{\sc ii-vi}, \\
          & O\,{\sc ii-vi}, Fe\,{\sc iii-vi} \\
        Metallicity & solar \citep{asplund09} \\
        \hline
    \end{tabular}
\end{table}

\addtolength{\tabcolsep}{-3pt}
\begin{table*}
    \centering
    \caption{Observed parameters for heavy metal stars, ordered by temperature (most recent measurement). The fourth column, $\log{y}$, is the logarithm of the ratio of helium to hydrogen number fractions. The abundances by number relative to solar are given in logarithmic form, $\log{\epsilon / \epsilon_{\odot}}$.}
    \label{tab:obs_params}
    \begin{tabular}{l|ccccccccc}
    \hline
        Star & $T_{\rm{eff}}$\,/\,kK & $\log{g/\mathrm{cm\,s^{-2}}}$ & $\log{y}$ & \multicolumn{5}{c}{$\log{\epsilon/\epsilon_{\odot}}$} & reference\\
         & & & & Ge & Sr & Y & Zr & Pb & \\
        \hline
         LS\,IV--14$^\circ$ 116 & 33$\pm$1 & 5.8$\pm$0.2 & --0.60$\pm$0.12 & - & - & - & - & - & \citet{viton91}\\

          & 35 & - & - & - & - & - & - & - & \citet{ulla98}\\
        
         & 32.50$\pm$0.15 & 5.4$\pm$0.1 & --0.58$\pm$0.02 & - & - & - & - & - & \citet{ahmad03a}\\
        
         & 34.0$\pm$0.5 & 5.6\,$\pm$\,0.2 & --0.72$\pm$0.10 & 2.87$\pm0.12$ & 3.92$\pm$0.10 & 3.99$\pm$0.15 & 3.93$\pm$0.24 & - & \citet{naslim11}\\

          & 34.95$\pm$0.25 & 5.93$\pm$0.04 & --0.62$\pm$0.03 & - & - & - & - & - & \citet{green11}\\
        
         & 35.150$\pm0.111$ & 5.88$\pm$0.02 & --0.62$\pm$0.01 & - & - & - & - & - & \citet{randall15}\\
         
        PHL 417 & 35.64$\pm$0.28 & 5.73$\pm$0.01 & --0.49$\pm$0.02 & 2.3$\pm$0.2 & - & 4.5$\pm$0.2 & 3.8$\pm$0.3 & - & \citet{ostensen20} \\
        
        Feige 46 & 36.12$\pm$0.23 & 5.93$\pm$0.04 & --0.32$\pm$0.03 & - & - & - & - & - & \citet{latour19a} \\
        
         & - & - & - & 2.43$\pm$0.63 & 4.61$\pm$0.41 & 4.67$\pm$0.41 & 4.31$\pm$0.11 & <2.83$^{+0.60}$ & \citet{latour19b} \\

         & - & - & - & 3.33$\pm$0.19 & 4.49$\pm$0.12 & 4.43$\pm$0.05 & 4.29$\pm$0.09 & <2.83$^{+0.60}$ & \citet{dorsch20} \\

        FBS 1749+373 & 34.63$\pm$0.6 & 5.89$\pm$0.12 & --0.28$\pm$0.06 & & & & & & \citet{nemeth12} \\
        
         & 36.8$\pm$2.0 & 5.80$\pm$0.20 & --0.43$\pm$0.14 & <1.67 & - & <3.26 & <3.45 & 3.14$\pm$0.16 & \citet{naslim20}\\

        HE 2359--2844 & 37.05$\pm$1.00 & 5.57$\pm$0.15 & --0.12$\pm$0.19 & - & - & 4.40$\pm$0.16 & 3.89$\pm$0.16 & 3.89$\pm$0.19 & \citet{naslim13}\\

        PG 1559+048 & 40.33$\pm$0.86 & 6.16$\pm$0.18 & --0.53$\pm$0.21 & & & & & & \citet{nemeth12} \\

         & 37.2$\pm$1.6 & 6.00$\pm$0.15 & --0.60$\pm$0.08 & 3.04$\pm$0.30 & - & 4.05$\pm$0.16 & <3.66 & 4.35$\pm$0.14 & \citet{naslim20} \\

        EC 22536--5304 & 35.55$\pm$1.50 & 5.92$\pm$0.15 & --0.70$\pm$0.05 & - & - & - & - & 4.78$\pm$0.25 & \citet{jeffery19b} \\

         & 38.0$\pm$0.4 & 5.81$\pm$0.04 & --0.15$\pm$0.04 & <2.50$^{+0.40}$ & <3.90$^{+0.40}$ & <4.07$^{+0.40}$ & <2.75$^{+0.40}$ & 6.27$\pm$0.32 & \citet{dorsch20} \\

        UVO 0825+15 & 38.90$\pm$0.27 & 5.97$\pm$0.11 & --0.57$\pm$0.01 & 2.59$\pm$0.06 & - & 3.16$\pm$0.09 & <2.72 & 3.74$\pm$0.18 & \citet{jeffery17a}\\

        HZ 44 & 39.1$\pm$0.6 & 5.64$\pm$0.10 & 0.08$\pm$0.05 & 2.13$^{+0.31}_{-0.34}$ & <3.71$^{+0.60}$ & <4.17$^{+0.20}$ & 3.18$^{+0.19}_{-0.20}$ & 4.04$^{+0.13}_{-0.14}$ & \citet{dorsch19}\\

        HE 1256--2738 & 39.5$\pm$1.0 & 5.66$\pm$0.10 & --0.02$\pm$0.17 & - & - & - & - & 4.64$\pm$0.25 & \citet{naslim13}\\

        HD 127493 & 42.48$\pm$0.25 & 5.60$\pm$0.05 & 0.60$\pm$0.30 & - & - & - & - & - & \citet{Hirsch09.thesis}\\

         & 42.07$\pm$0.18 & 5.61$\pm$0.04 & 0.33$\pm$0.06 & 2.67$^{+0.40}_{-0.44}$ & <3.55$^{+0.30}$ & <4.41$^{+0.30}$ & <3.44$^{+0.20}$ & 3.92$^{+0.40}_{-0.44}$ & \citet{dorsch19}\\
        \hline
    \end{tabular}
\end{table*}
\addtolength{\tabcolsep}{2pt}

\begin{figure}
    \centering
    \includegraphics[width=\linewidth]{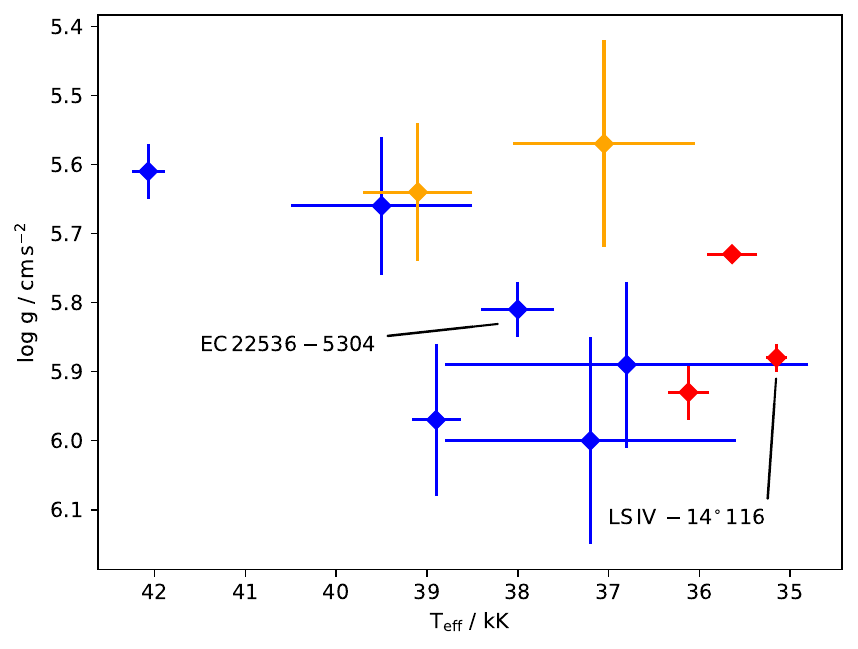}
    \caption{Effective temperatures and surface gravities for the heavy metal stars listed in Table\,\ref{tab:obs_params} (most recent measurements). Markers for each star are coloured according to their abundances of Zr and Pb. Stars enriched with Zr are in red, those enriched with both Zr and Pb are yellow, and those enriched with Pb are blue. The stars mentioned in the text are labelled individually.}
    \label{fig:gt}
\end{figure}

This study used the stellar atmosphere modelling code {\sc tlusty} version 208 and the spectral synthesis code {\sc synspec} version 54 \citep{hubeny17a,hubeny17b,hubeny17c,hubeny21} to model the atmospheres and spectra of stars with a stratified abundance of lead. {\sc synspec} was modified similarly to \citet{latour19b} and \citet{dorsch19} to include two-, three- and four-times-ionised lead (Pb\,{\sc iii}, Pb\,{\sc iv} and Pb\,{\sc v}). As these ions are not included in the default {\sc synspec} linelist, we used the linelist of \citet{dorsch19}. Atomic data for lead lines included in this linelist were compiled from \citet{alonso09}, \citet{morton00}, \citet{naslim13}, \citet{safranova04}, \citet{alonso11}, \citet{ansbacher88} and \citet{colon14}.

The {\sc tlusty} models used in this study are summarised in Table\,\ref{tab:model_params}. To simulate the observed parameter range of heavy metal subdwarfs, the atmospheric models were calculated at 34, 38 and 42\,kK. For reference, observed parameters of heavy metal subdwarfs are summarised in Table\,\ref{tab:obs_params} and Fig.\,\ref{fig:gt}. All models have $\log{g/\mathrm{cm\,s^{-2}}}=5.80$ and are composed of 50 per cent H by number, with the remaining 50 per cent being made of He and solar metal abundances. Opacity contributions came from H, He, C, N, O and Fe only, in non-local thermodynamic equilibrium (non-LTE). The model atoms were those distributed with version 208 of {\sc tlusty}, and can be found on the {\sc tlusty} website\footnote{http://tlusty.oca.eu/Tlusty2002/tlusty-frames-data.html}. These model atoms used level energies from the NIST database \citep{ralchenko20} and photoionisation cross sections from the Opacity Project database \citet{cunto93}. The {\sc tlusty} models had 50 depth points, spanning a range in log column mass from -6.4 to 1.6 (units ${\rm  g\,cm}^{-2}$). Lead was treated in LTE due to the lack of photoionisation cross-sections necessary for non-LTE.

For each {\sc tlusty} model, a set of {\sc synspec} model spectra were calculated for different lead stratifications. Only vertical stratification was considered in this study, which was approximated by a box function with an abundance $A_{\rm{Pb}}$, chosen to be 4 dex above solar. Figure\,\ref{fig:sketch} illustrates the size and position of this lead layer in terms of column mass (g\,cm$^{-2}$), with the top and the bottom of the layer given by \mzero{} and \mone{} respectively. The width and position of the lead layer was varied over all possible depth point combinations, with the widest layer encompassing the whole 50 depth points and the narrowest layers being enriched in only two depth points, those being \mzero{} and \mone{}.

\begin{table*}
\caption{Lines used in the study, with the wavelength, $\lambda$, the oscillator strength, $\log{gf}$, and the energy of the lower level with respect to the ground state of the ion, $E_{\rm l}$. Lines are ordered first by ion, shown at the top of each subtable, then by wavelength. These quantities are from the linelist of \citet{dorsch19}. Lines marked `*' were measured as part of a blend with the wavelength above. The line marked `$\dagger$' was measured as part of a blend with the Pb\,{\sc iv} lines at 1313\,\AA{}.}
\label{tab:lines}
\begin{subtable}{0.2\linewidth}
\centering
\begin{tabular}[t]{lcc}
    $\lambda$ / \AA & $\log{gf}$ & $E_{\rm l}$ / cm$^{-1}$\\
    \hline
    Pb\,{\sc iii} \\
	1005.500 & -0.201 & 78984.6 \\
	1030.400 & -0.027 & 60397.0 \\
	1041.000 & -0.180 & 150083.7 \\
	1048.870 & 0.114 & 0.0 \\
	1052.200 & -0.367 & 60397.0 \\
	1069.100 & 0.317 & 64391.0 \\
	1074.600 & -0.205 & 64391.0 \\
	1098.400 & -0.541 & 64391.0 \\
	1115.000 & -0.640 & 60397.0 \\
	1118.700 & -1.008 & 64391.0 \\
	1147.000 & -0.587 & 158956.8 \\
	1167.000 & -0.213 & 64391.0 \\
	1176.000 & -0.440 & 142551.0 \\
	1203.500 & 0.526 & 95340.1 \\
	1216.700 & -0.020 & 142551.0 \\
	1250.400 & 0.507 & 78984.6 \\
	1266.800 & -0.347 & 78984.6 \\
	1274.500 & -1.429 & 78984.6 \\
	1308.100 & -0.361 & 78984.6 \\
	$^{\dagger}$1313.100 & -0.146 & 151884.5 \\
	1406.500 & -0.041 & 78984.6 \\
	1426.200 & -0.318 & 157925.0 \\
	1447.500 & -0.121 & 158956.8 \\
	1486.000 & -0.485 & 157444.1 \\
	$^*$1486.000 & -0.015 & 155431.5 \\
	1496.700 & -1.152 & 157925.0 \\
	1553.021 & -1.135 & 0.0 \\
	1587.600 & 0.439 & 164817.9 \\
	1597.800 & -1.065 & 95340.1 \\
	1610.200 & -1.444 & 95340.1 \\
	1664.100 & -1.790 & 95340.1 \\
	1668.800 & -1.268 & 164817.9 \\
	1711.100 & -0.341 & 95340.1 \\
\end{tabular}
\end{subtable}
\hfill
\begin{subtable}{0.2\linewidth}
\centering
\begin{tabular}[t]{lcc}
    $\lambda$ / \AA & $\log{gf}$ & $E_{\rm l}$ / cm$^{-1}$\\
    \hline
    Pb\,{\sc iii}\\
    1826.700 & -1.481 & 95340.1 \\
    3044.524 & 0.498 & 157444.1 \\
	3138.524 & 0.635 & 157925.0 \\
	3177.224 & 0.815 & 158956.8 \\
    3531.200 & -0.707 & 171081.4 \\
	3656.538 & 0.724 & 190287.8 \\
	3672.239 & 0.959 & 190429.0 \\
	3689.310 & -0.426 & 150083.7 \\
	3706.941 & -0.429 & 170917.3 \\
	3729.542 & -0.080 & 171081.4 \\
	3736.843 & 0.870 & 190901.2 \\
	3854.080 & 0.302 & 150083.7 \\
	3951.920 & -1.284 & 151884.5 \\
	4031.160 & -0.429 & 174600.9 \\
	4272.660 & -0.462 & 153783.4 \\
	4499.340 & -0.666 & 177181.4 \\
	4571.219 & 0.029 & 176022.9 \\
	4761.120 & 0.012 & 150083.7 \\
	4798.590 & -0.356 & 150083.7 \\
	4828.188 & -0.749 & 177181.4 \\
	5003.400 & -0.694 & 177906.5 \\
	5065.120 & -1.029 & 157444.1 \\
	5191.560 & -0.594 & 157925.0 \\
	5523.970 & -0.641 & 157925.0 \\
	5779.410 & -0.745 & 153783.4 \\
	5857.960 & 0.189 & 158956.8 \\
	6608.300 & -0.842 & 184268.1 \\
	7422.100 & -0.406 & 157444.1 \\
	7600.900 & -0.187 & 157925.0 \\
	7675.496 & 0.635 & 201398.7 \\
	7768.670 & 0.782 & 201597.3 \\
	8046.267 & 0.942 & 202046.8 \\
	8725.625 & 0.842 & 203301.6 \\
\end{tabular}
\end{subtable}
\hfill
\begin{subtable}{0.2\linewidth}
\centering
    \begin{tabular}[t]{lcc}
    $\lambda$ / \AA & $\log{gf}$ & $E_{\rm l}$ / cm$^{-1}$ \\
    \hline
    Pb\,{\sc iv} \\
	1028.611 & 0.056 & 0.0 \\
	1032.050 & 0.394 & 122567.8 \\
	1116.090 & 0.528 & 97218.5 \\
	1144.940 & -0.428 & 97218.5 \\
	1313.027 & -2.042 & 0.0 \\
	$^*$1313.053 & -3.664 & 0.0 \\
	$^*$1313.061 & -1.434 & 0.0 \\
	$^*$1313.070 & -0.613 & 0.0 \\
	$^*$1313.073 & -1.740 & 0.0 \\
	$^*$1313.084 & -2.042 & 0.0 \\
	1798.400 & 0.017 & 214891.7 \\
	1959.400 & 0.708 & 219463.4 \\
	2462.229 & 0.371 & 209789.4 \\
	2509.681 & -0.281 & 209789.4 \\
	2865.132 & 0.429 & 184559.3 \\
	$^*$2865.371 & 0.563 & 186817.0 \\
	2979.063 & 0.646 & 217852.6 \\
	3003.431 & -0.725 & 184559.3 \\
	3052.560 & 0.317 & 185103.5 \\
	3063.108 & -0.736 & 186817.0 \\
	3071.983 & -0.280 & 217852.6 \\
	3221.170 & 0.237 & 186817.0 \\
	3231.751 & -0.277 & 219461.7 \\
	3366.411 & -0.178 & 221716.5 \\
	3962.467 & -0.047 & 184559.3 \\
	4049.832 & -0.065 & 185103.5 \\
	4174.478 & -0.444 & 185103.5 \\
	4496.223 & -0.437 & 186798.0 \\
	4534.447 & 1.190 & 270496.7 \\
	$^*$4534.917 & 1.102 & 270499.5 \\
    4605.400 & -0.991 & 186817.0 \\
	5914.540 & -0.907 & 184559.3 \\
	6113.600 & -1.337 & 185103.5 \\
    \end{tabular}
\end{subtable}
\hfill
\begin{subtable}{0.2\linewidth}
\centering
    \begin{tabular}[t]{lcc}
    $\lambda$ / \AA & $\log{gf}$ & $E_{\rm l}$ / cm$^{-1}$ \\
    \hline
    Pb\,{\sc v} \\
	1051.246 & -0.887 & 132714.8 \\
	1064.300 & -1.430 & 245279.8 \\
	1088.860 & -0.306 & 136001.1 \\
	1096.523 & -0.866 & 132714.8 \\
	1104.838 & -1.128 & 136001.1 \\
	1137.500 & -0.765 & 136001.1 \\
	1152.363 & -2.604 & 132714.8 \\
	1157.879 & -0.106 & 110770.2 \\
	1183.700 & -1.143 & 245279.8 \\
	1185.441 & -0.081 & 132714.8 \\
	1189.953 & 0.350 & 110770.2 \\
	1197.700 & -0.394 & 136001.1 \\
	1213.200 & 0.147 & 114708.5 \\
	1233.500 & -0.316 & 136001.1 \\
	1248.460 & -1.046 & 114708.5 \\
	1610.506 & -2.907 & 132714.8 \\
	1635.700 & -1.783 & 136001.1 \\
	1700.500 & -1.833 & 136001.1 \\
	2140.100 & -0.257 & 329578.1 \\
	2222.200 & 0.041 & 328238.5 \\
	2267.400 & -0.316 & 332505.2 \\
	2324.200 & 0.115 & 332505.2 \\
	2641.000 & 0.617 & 338445.4 \\
    \end{tabular}
\end{subtable}
\end{table*}

\section{Results}
\label{sec:results}

\subsection{Stratification effects on the equivalent width}
\label{sec:cusp}
For each enriched layer configuration, the profiles of the lead lines listed in Table\,\ref{tab:lines} were calculated in a 2\,\AA{} region centred on each line. The equivalent width was then calculated for each line, enriched layer and temperature combination.

\begin{figure}
    \centering
    \includegraphics[width=\columnwidth]{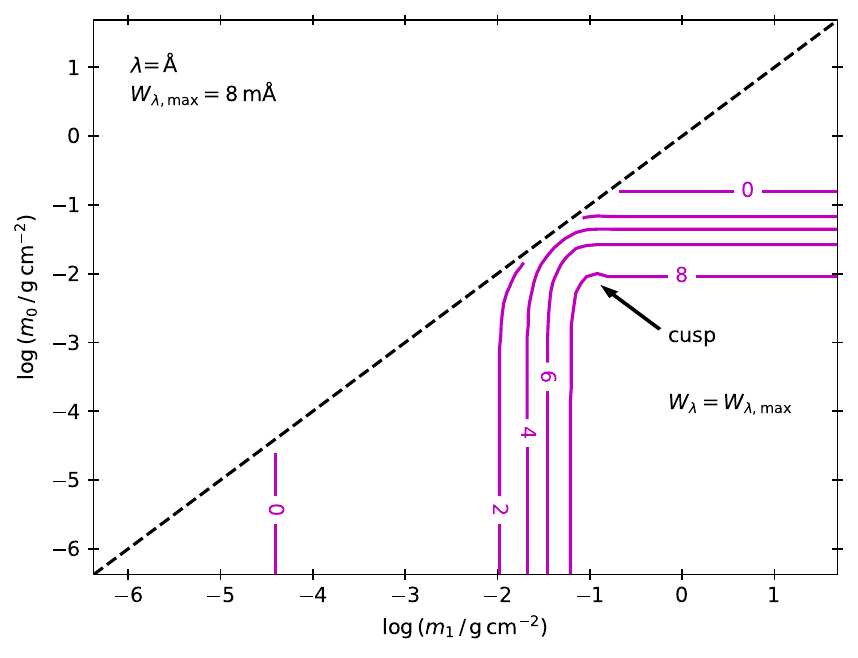}
    \caption{Contours of equivalent width in m\AA{} in the $m_{\rm 0}$-$m_{\rm 1}$ plane for the Pb\,{\sc iv} line at 4049\,\AA{} for $T_{\rm{eff}}=38$\,kK. The point corresponding to the narrowest enriched layer which still results in maximum equivalent width is labelled as `cusp'. Visualisations of the stratification for different regions of the plane can be seen on the right-hand panel of Fig.\,\ref{fig:sketch}.}
    \label{fig:cusp_example}
\end{figure}

The enriched layer size and position affect both the line strength, measured by equivalent width $W_{\rm \lambda}$, and line shape (i.e. broad line profile versus narrow, deep versus shallow). The effects on line strength can be visualised with contours of $W_{\lambda}$ on the \mzero-\mone{} plane. This is shown in Fig.\,\ref{fig:cusp_example} for the Pb\,{\sc iv} line at 4049\,\AA{}, which is the strongest optical line in our study. Per definition, all of the combinations of \mzero{} and \mone{} for which model line profiles were calculated lie below the dashed black line at \mzero=\mone.

The contours on Fig.\,\ref{fig:cusp_example} show how equivalent width changes with stratification. The point labelled `cusp' on Fig.\,\ref{fig:cusp_example} corresponds to the narrowest enriched layer which still produces maximum equivalent width. The equivalent width of all lines shows a qualitatively similar behaviour in the \mzero-\mone{} plane, with equivalent width being maximum at the cusp and all points below and to the right of it.

\begin{figure*}
    \centering
    \includegraphics[width=0.9\linewidth]{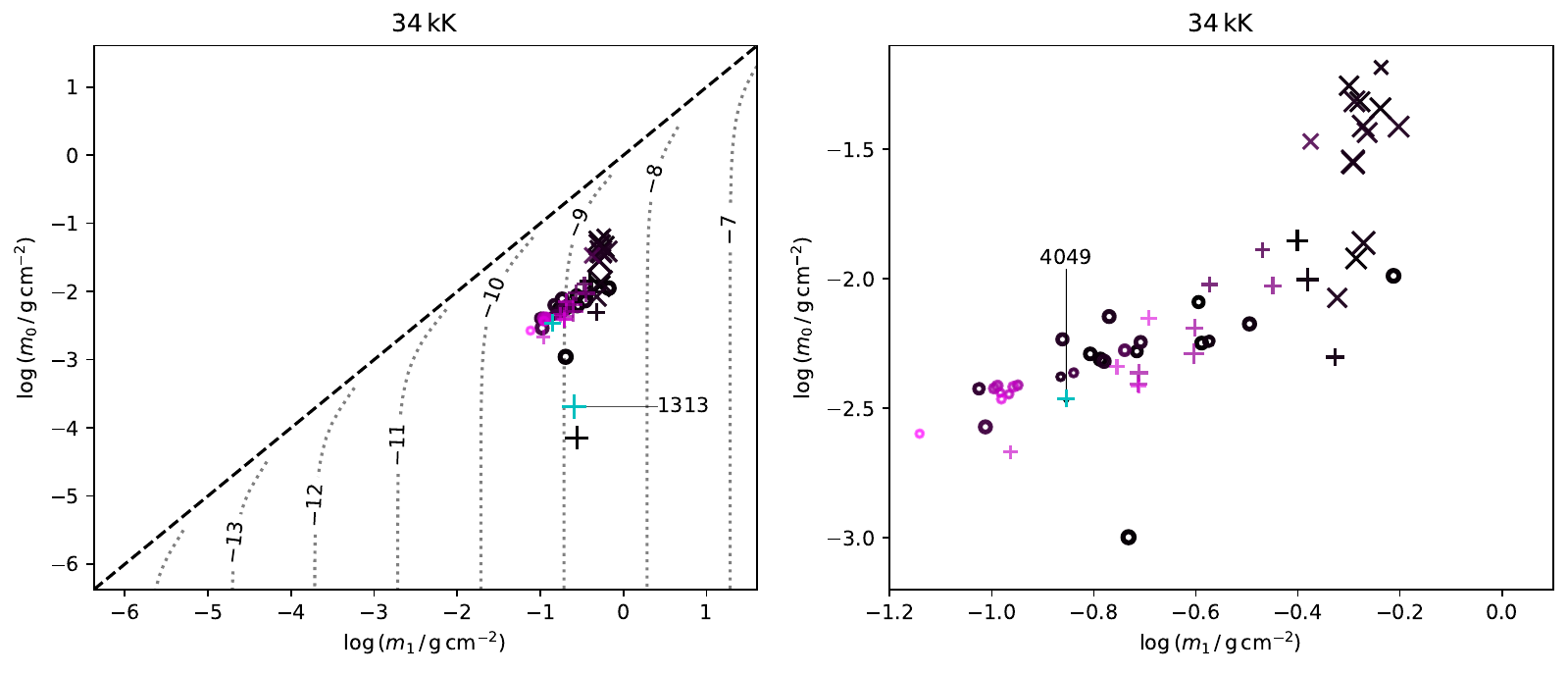} 
    \includegraphics[width=0.9\linewidth]{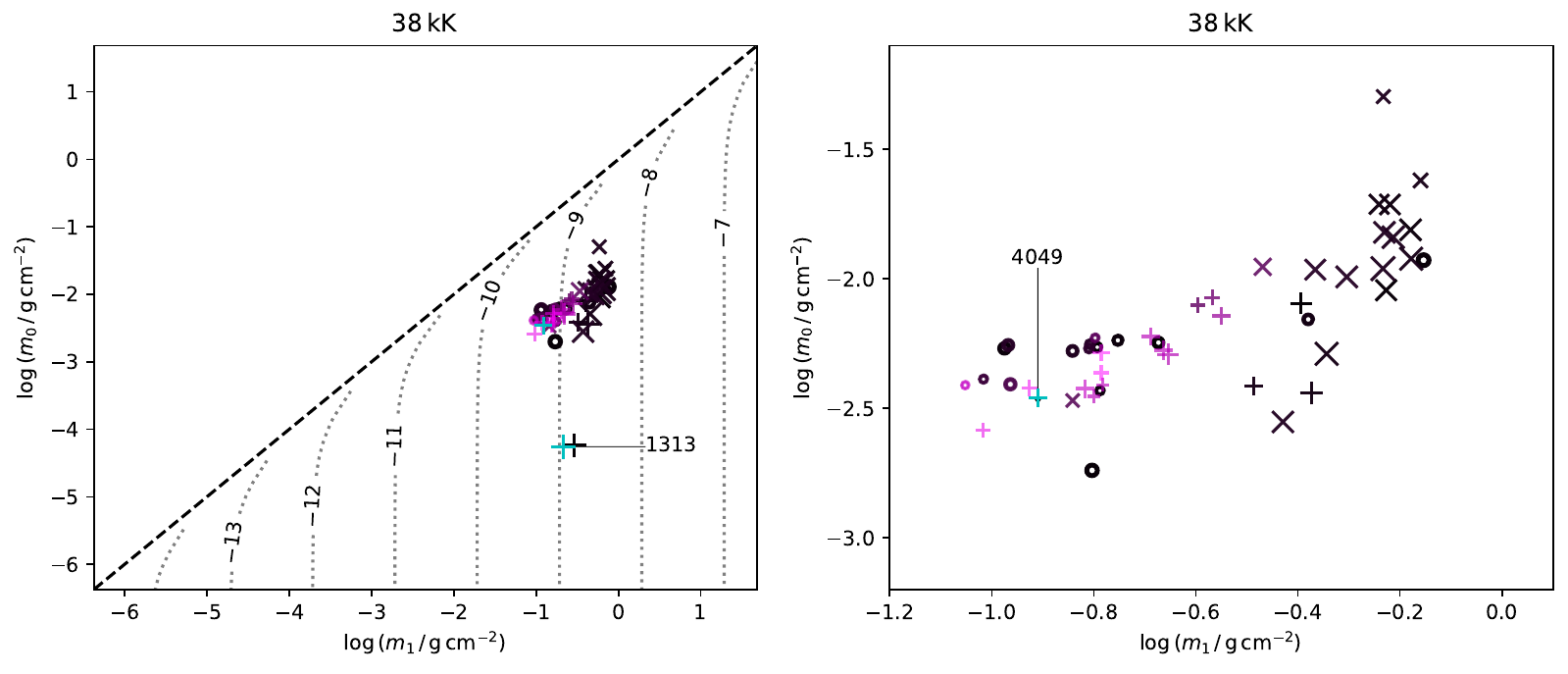}
    \includegraphics[width=0.9\linewidth]{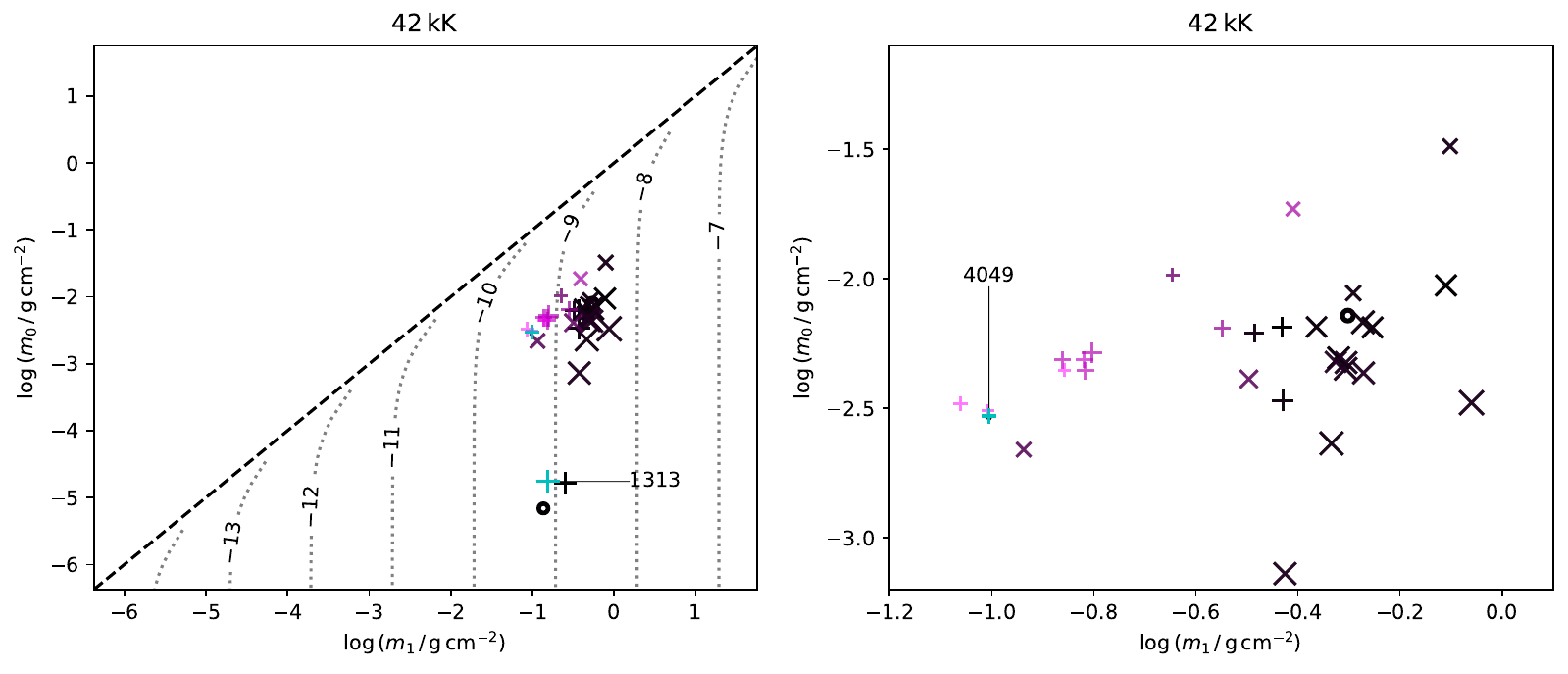}
    \caption{Cusp position for all lines with maximum equivalent width greater than 2\,m\AA, with the colour gradient indicating wavelength (from shortest wavelength in black to longest wavelength in magenta). The left-hand panels show the entire \mzero{}-\mone{} plane, whereas the right-hand panels show a zoom-in on the main cloud of cusp positions. The cyan markers show the cusps of the lines at 1313 and 4049\,\AA{}, also shown in Fig.\,\ref{fig:line_profiles}. The point size scales with the maximum equivalent width of the line. Ion is indicated by marker type - `o' for Pb\,{\sc iii}, `+' for Pb\,{\sc iv}, and `x' for Pb\,{\sc v}. Some jitter has been added to reduce overlap between points. The contours are the logarithm of $M_{\mathrm{mixed}}$, given by Eq.\,\ref{eq:mass}.}
    \label{fig:spot}
\end{figure*}

\begin{figure}
    \centering
    \includegraphics[width=\columnwidth]{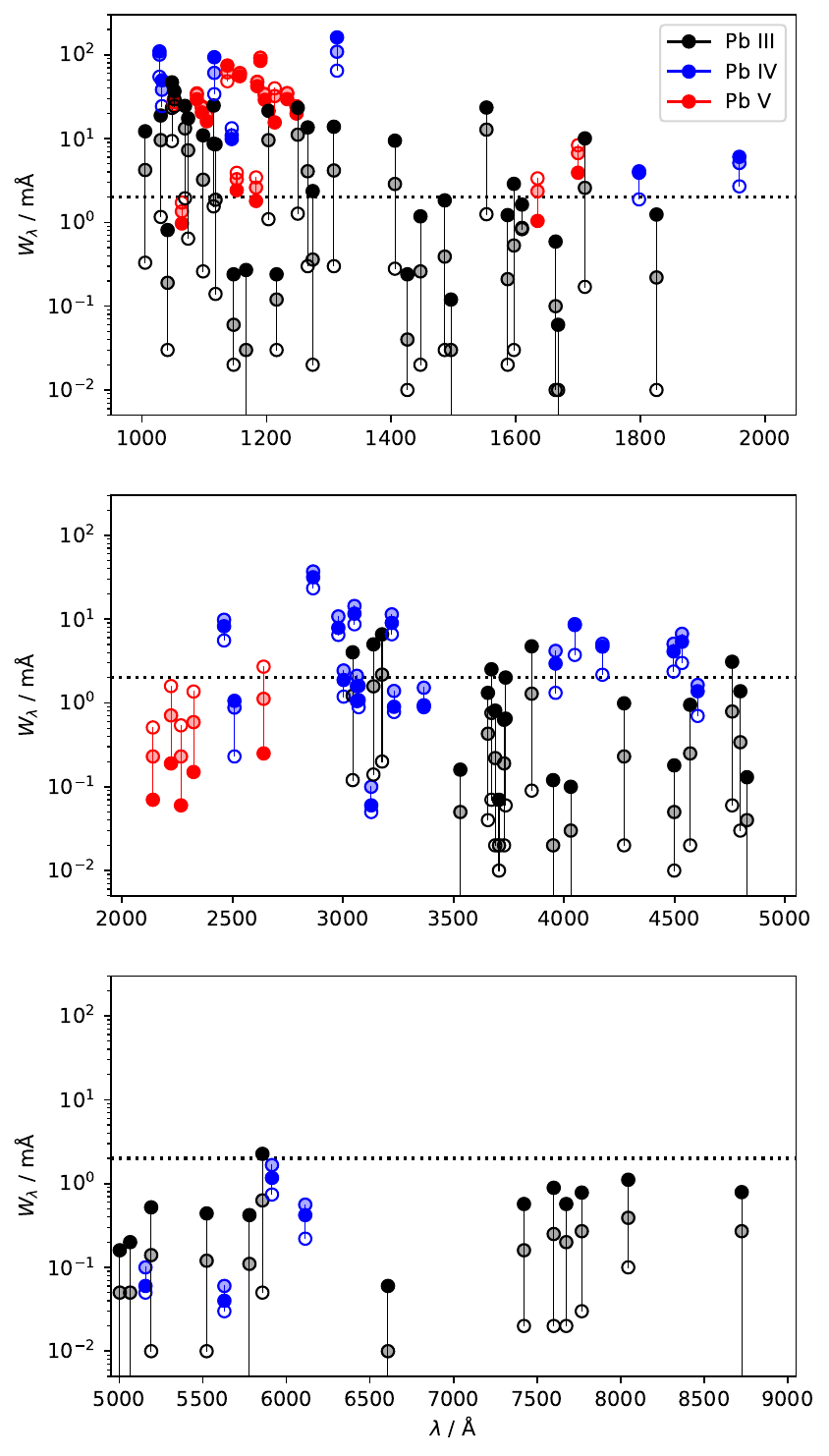}
    \caption{Maximum equivalent width against wavelength for all lines in the study, and at all three temperatures. Temperature is indicated by the transparency of the point, with 34, 38 and 42\,kK indicated by solid, shaded and open markers respectively. The dotted horizontal line shows the 2\,m\AA{} cutoff for the inclusion of lines in Figs.\,\ref{fig:spot} and \ref{fig:stripe}. Note that the span in wavelength is different for each panel.}
    \label{fig:maxew}
\end{figure}

However, the position of the cusp is slightly different for each line. Figure\,\ref{fig:spot} demonstrates this by showing the approximate cusp position, at 97 per cent of the maximum equivalent width\footnote{The plateau of maximum equivalent width, as seen in Fig.\,\ref{fig:cusp_example}, actually has a shallow slope. Taking 97 per cent of the maximum equivalent width rather than 100 per cent for the calculation of the cusp position excludes this slope.}, for all lines with a maximum equivalent width greater than 2\,m\AA. Wavelength trends can be identified by the colour gradient of the markers on Fig.\,\ref{fig:spot}, from longer wavelengths in magenta to shorter wavelengths in black. The lines which exceed the 2\,m\AA{} cutoff can be seen in Fig.\,\ref{fig:maxew}, which shows how the maximum equivalent width for each line changes with temperature.

In Fig.\,\ref{fig:spot}, the cusp positions for the model atmospheres at 34 and 38\,kK are similar, indicating that the minimal layer required for maximum equivalent width does not change for most Pb lines in this temperature range. For all three temperatures, longer wavelengths tend towards a cusp with lower \mone{}; that is, longer wavelength lines can reach maximum equivalent width without needing the lead layer to be as deep, and are therefore most insensitive to the position of \mone{}. The Pb\,{\sc v} lines, which lie at shorter wavelengths, generally have cusp positions with higher \mone{}, meaning that they are more sensitive to the position of the bottom of the lead layer.

There are also a few lines below the main cloud of cusp positions. For all temperatures, these include the resonance lines of Pb\,{\sc iv}. For 42\,kK, the Pb\,{\sc iii} 1048\,\AA{} line also lies in this region. Being some of the strongest lines in the study, these lines have significant, although not maximum, equivalent width in some regions of the plane with deeper \mzero{} or shallower \mone{} than the cusp. The low \mzero{} of their cusps means that the profiles of the resonance lines are more sensitive to the position of the upper boundary of the lead layer than other lines.

The mass of lead contained in the enriched layer varies across the \mzero-\mone{} plane. Assuming that the enriched layer of lead is formed by diffusion alone, an important question is how much mass of normal (e.g. solar) composition contains the same amount of lead as the enriched layer. This quantity, $M_{\mathrm{mixed}}$,  represents the amount of mass over which diffusion must be active to concentrate the necessary lead abundance and hence explain the observed lines.

In order to estimate $M_{\mathrm{mixed}}$, we assume that the atmosphere initially has a homogeneous lead mass fraction, $X_{\mathrm{Pb,ini}}$. After transport of lead by diffusion, the mass fraction of lead within the layer is $X_{\mathrm{Pb,layer}}$. Assuming that all the lead within $M_{\mathrm{mixed}}$ can be concentrated into the the layer between \mzero{} and \mone{}, we then have  
\begin{equation}
    \frac{M_{\mathrm{mixed}}}{M_{\mathrm{layer}}} = \frac{X_{\mathrm{Pb,layer}}}{X_{\mathrm{Pb,ini}}},
\end{equation}
where $M_{\mathrm{layer}}$ is the mass contained between \mzero{} and \mone{}. Since the layer is thin compared to the stellar radius, $M_{\mathrm{layer}}$ can be estimated by multiplying the enclosed column mass by the surface area of the star. This then gives

\begin{equation}
    \label{eq:mass}
    \begin{split}
    M_{\mathrm{mixed}} & \approx \frac{X_{\mathrm{Pb,layer}}}{X_{\mathrm{Pb,ini}}} M_{\mathrm{layer}}\\
    & \approx \frac{X_{\mathrm{Pb,layer}}}{X_{\mathrm{Pb,ini}}} 4\pi R_*^2 (m_{\rm 1} - m_{\rm 0}),
    \end{split}
\end{equation}

where $R_*$ is the radius of the star. The dotted grey contours in Fig.\,\ref{fig:spot} show the logarithm of $M_{\mathrm{mixed}}$ in solar masses, assuming $R_*=0.13$\,R$_{\odot}$, $X_{\mathrm{Pb,layer}}=10^4 X_{\mathrm{Pb,ini}}$ and a solar initial lead abundance from \citet{asplund09}. Most cusp positions lie between $10^{-8}$ and $10^{-9}$\,M$_{\rm \odot}$, with a few lines requiring less mixed mass. Note that EC\,22536--5304, for which $X_{\mathrm{Pb,layer}}=10^8 X_{\mathrm{Pb,ini}}$ \citep{dorsch21}, would require $10^4$ times as much mixed mass to produce an enriched layer in the same position as this example.

\begin{figure*}
    \centering
    \includegraphics[width=0.9\linewidth]{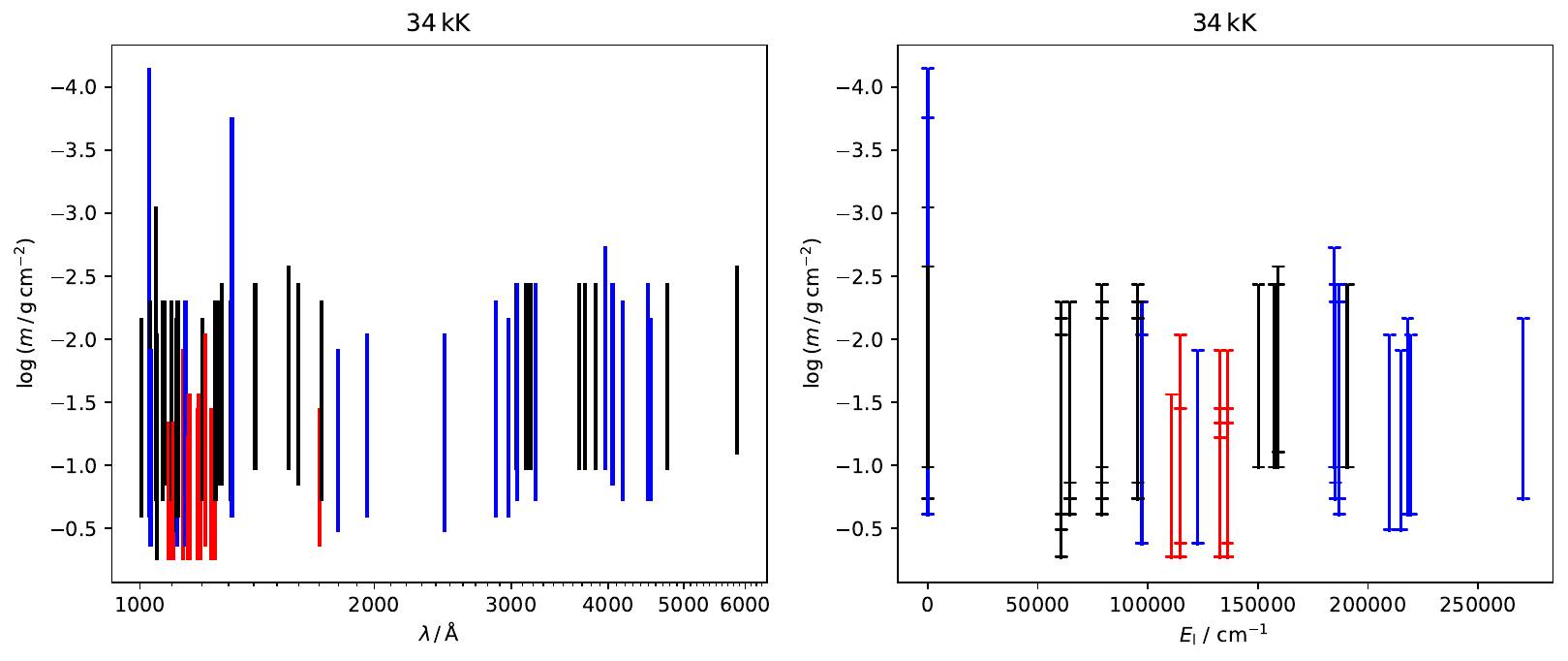} 
    \includegraphics[width=0.9\linewidth]{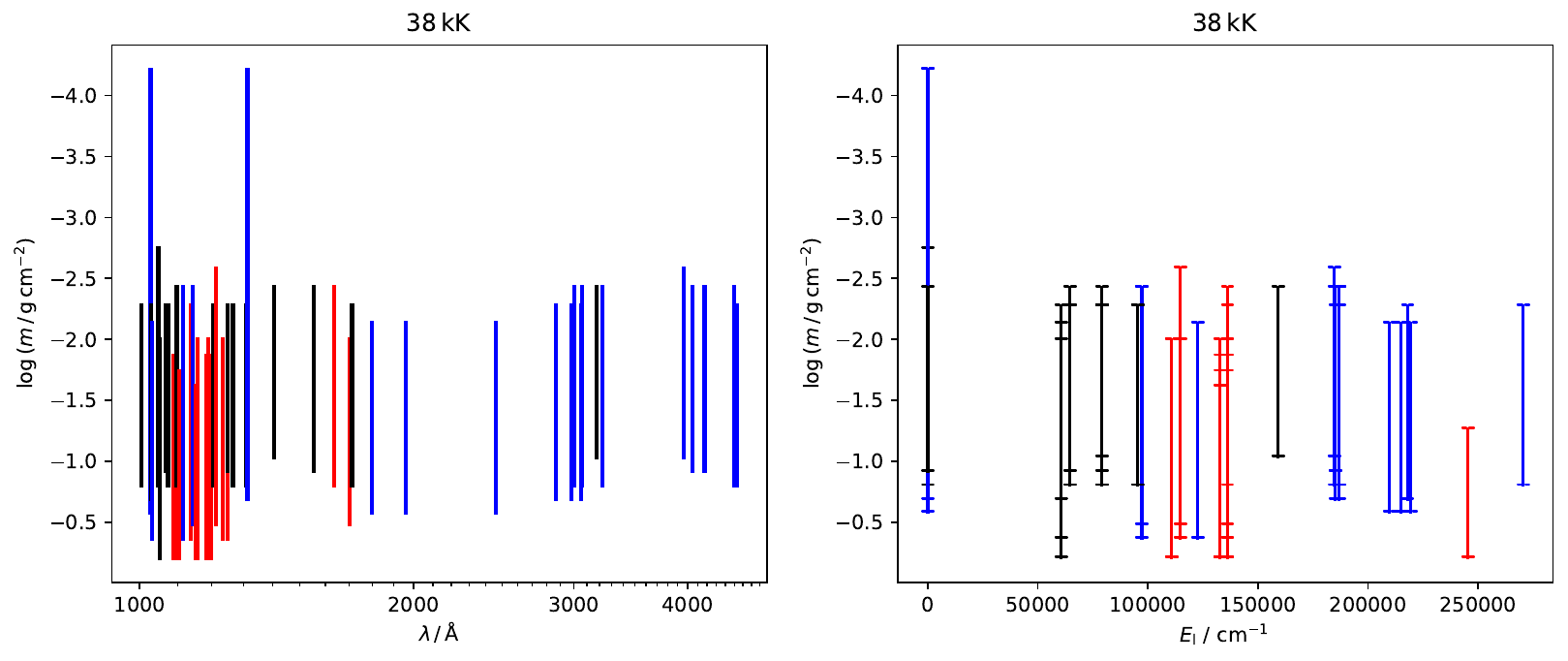}
    \includegraphics[width=0.9\linewidth]{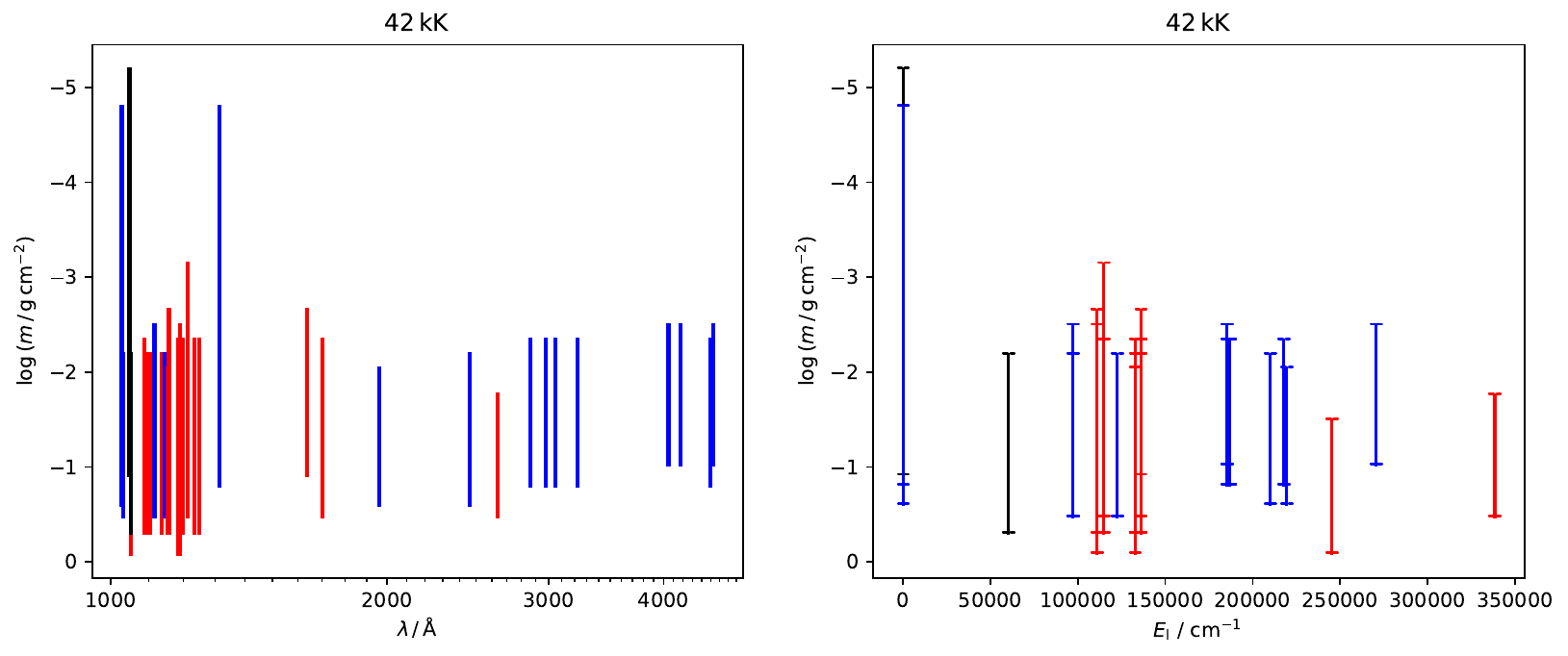}
    \caption{Log column mass encompassed by the enriched layer at the cusp against wavelength (left-hand panels) and energy of the lower level (right-hand panels) for each effective temperature. The top of each bar is at \mzero{} and the bottom is at \mone{}. Note that horizontal caps on each line show where lines overlap in the right-hand panels. The bar colour corresponds to ion, with black, blue and red referring to Pb\,{\sc iii}, {\sc iv} and {\sc v} respectively. As in Fig.\,\ref{fig:spot}, only lines with maximum equivalent width > 2\,m\AA{} are shown.}
    \label{fig:stripe}
\end{figure*}

To demonstrate the cusp position more clearly, the column mass included in the enriched layer at the cusp for lines with maximum equivalent width above 2\,m\AA{} is shown in the left-hand panels of Fig.\,\ref{fig:stripe}. The top of each bar is at $m_{\rm 0}$ and the bottom is at $m_{\rm 1}$. In the right-hand panels is the cusp position versus the energy of the lower level, for which there seems to be no discernible dependence. The Pb\,{\sc iv} (and in the case of the higher effective temperatures, Pb\,{\sc iii}) resonance lines have the largest span in \mzero{} and \mone{}, and are therefore most sensitive to stratification, particularly if log(\mzero) lies between -3 and -5. The lower boundary of the enriched layer is best probed by the Pb\,{\sc v} lines in the UV.

\subsection{Effects of stratification on line profiles}
\label{sec:profiles}
\begin{figure}
    \centering
    \includegraphics[width=\columnwidth]{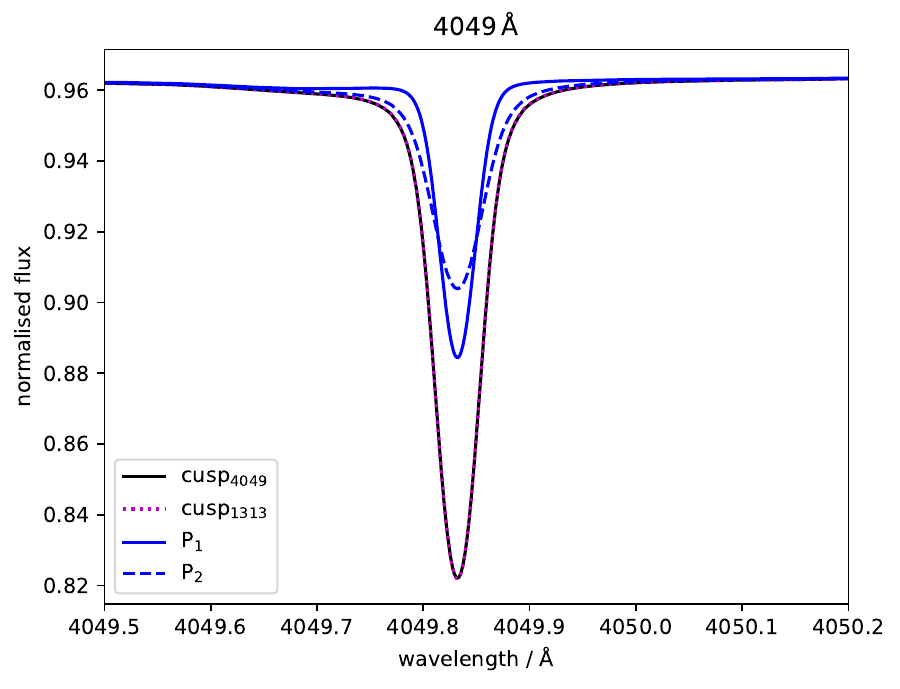}\\
    \includegraphics[width=\columnwidth]{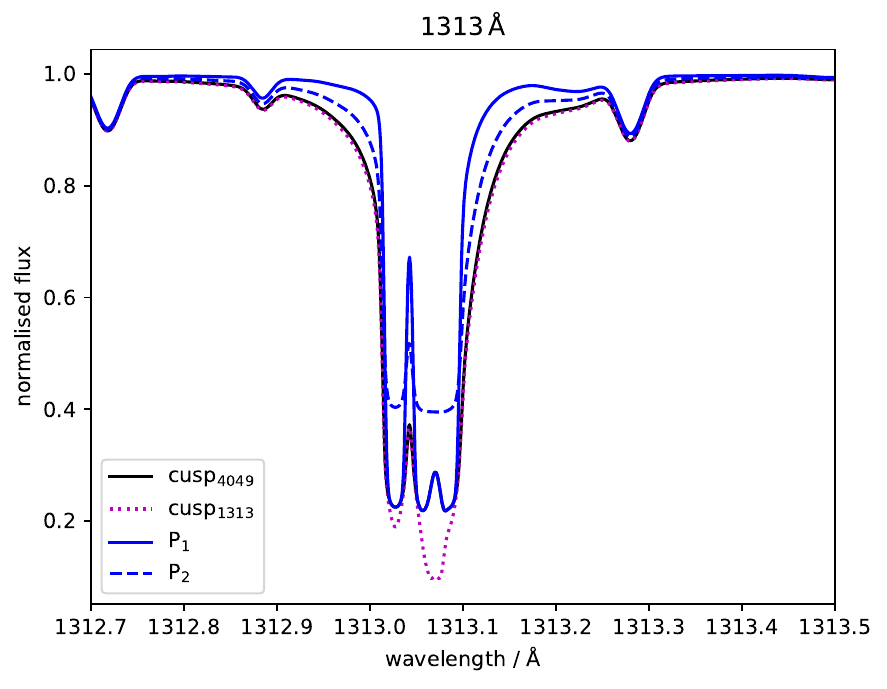}\\
    \includegraphics[width=\columnwidth]{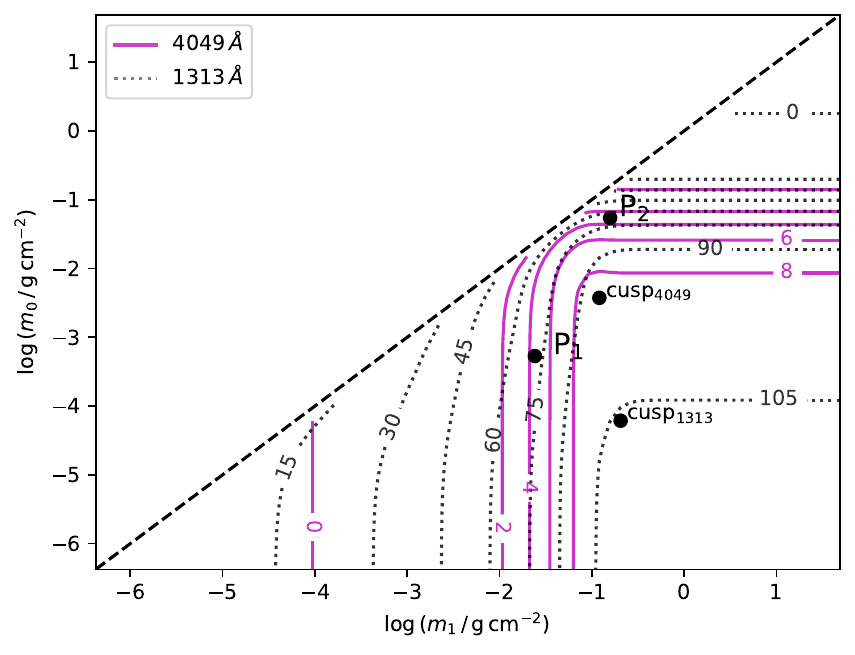}\\
    \caption{Line profiles at 4049\,\AA{} (top) and 1313\,\AA{} (middle) at $T_{\rm{eff}}=38\,$kK, with the configurations of the corresponding lead-enriched layers (bottom). The bottom panel also includes contours of equivalent width for both lines, labelled in m\AA{}. Points labelled `cusp' occur at the narrowest enriched layer configuration with maximum equivalent width for each line. The points P$_{\rm 1}$ and P$_{\rm 2}$ show the positions of two example lead layers narrower than at the cusps. P$_{\rm 1}$ and P$_{\rm 2}$ have been selected to produce the same equivalent width as one another, considering each spectral line separately ($\sim$5\,m\AA{} for 4049\,\AA{}, $\sim$70\,m\AA{} for 1313\,\AA{}). P$_{\rm 2}$ is deeper in the atmosphere than P$_{\rm 1}$.}
    \label{fig:line_profiles}
\end{figure}

As shown in Fig.\,\ref{fig:spot} and Fig.\,\ref{fig:stripe}, the effect on equivalent width of varying the size and position of the lead layer is wavelength dependent. Therefore, lines with a large enough difference in wavelength can be expected to behave differently in a certain range of \mzero{} and \mone. To demonstrate this, Fig.\,\ref{fig:line_profiles} shows line profiles corresponding to four selected points in the \mzero-\mone{} plane for the strong optical line at 4049\,\AA{} and the Pb\,{\sc iv} resonance line at 1313\,\AA. These points are cusp$_{\rm 4049}$ and cusp$_{\rm 1313}$ (the cusp points of their respective lines), and P$_{\rm 1}$ and P$_{\rm 2}$. The profiles of the cusp points in Fig.\,\ref{fig:line_profiles} show how the lines appear at maximum equivalent width. Points P$_{\rm 1}$ and P$_{\rm 2}$ show the appearance of lines formed in shallow and deep layers respectively. P$_{\rm 1}$ and P$_{\rm 2}$ were chosen such that at 4049\,\AA{}, P$_{\rm 1}$ and P$_{\rm 2}$ both produce lines with $\sim$5\,m\AA{} equivalent width, whereas at 1313\,\AA{} they both produce $\sim$70\,m\AA{} equivalent width. The location of each of these points in the \mzero{}-\mone{} plane is indicated in the lower panel of Fig.\,\ref{fig:line_profiles}.

Firstly, the 4049\,\AA{} line has the same profile at both cusp$_{\rm 4049}$ and cusp$_{\rm 1313}$. The line profile does not change because the lead layer at cusp$_{\rm 1313}$ extends further both upwards and downwards into the atmosphere than the layer at cusp$_{\rm 4049}$. Since maximum equivalent width is already reached at cusp$_{\rm 4049}$, this extension has no further effect on the profile.

The 4049\,\AA{} line profiles at P$_{\rm 1}$ and P$_{\rm 2}$ have both less equivalent width and a different line shape compared to that seen at cusp$_{\rm 4049}$. Of these two points, the lead layer of P$_{\rm 2}$ lies deeper in the atmosphere. It thus has a broadened shape compared to the profile at P$_{\rm 1}$, as the absorption occurs at a higher temperature and pressure. Conversely, the profile at P$_{\rm 1}$ is narrower with more absorption in the centre, as its lead layer is located in a cooler, less dense part of the atmosphere, and is subject to less broadening. This is similar to ordinary, non-stratified lines, whose cores are formed in shallower layers than the more broadened wings.

The 1313\,\AA{} line shows similar behaviour at P$_{\rm 1}$ and P$_{\rm 2}$, with P$_{\rm 1}$ showing stronger cores and P$_{\rm 2}$ showing stronger wings. Note that there are multiple lead lines at 1313\,\AA{}, so the overall profile is a blend. The equivalent width of the 1313\,\AA{} line is not at maximum at cusp$_{\rm 4049}$. The largest difference between cusp$_{\rm 1313}$ and cusp$_{\rm 4049}$ is that \mzero{} is lower at cusp$_{\rm 1313}$. This means the lead layer extends into shallower parts of the atmosphere at cusp$_{\rm 1313}$, so the equivalent width difference occurs mainly in the line core.

The line component with the highest $\log{gf}$ is at 1313.07\,\AA{}. This component therefore reaches saturation at a higher temperature than the others, hence the unusual line core profiles for P$_{\rm 1}$ and cusp$_{\rm 4049}$, where this strong component does not emerge. At cusp$_{\rm 1313}$, the other line components begin to experience saturation effects and the 1313.07\,\AA{} component appears again. At the abundance used in our models, these saturation effects cause the lines at 1313\,\AA{} to be sensitive to the abundance profile (see Fig.\,\ref{fig:gauss_1313}).

\subsection{Effects of temperature and ionisation fraction}
\label{sec:ionisation}

As expected, the lead line strengths differ between models of different effective temperature, with higher ionisation stages being generally stronger in the hotter models. As previously seen in Fig.\,\ref{fig:maxew}, the Pb\,{\sc iii} lines are generally strongest at 34\,kK, and Pb\,{\sc v} lines are strongest at 42\,kK. There are exceptions, e.g. the low-lying Pb\,{\sc v} lines $\sim$1200\,\AA.

\begin{figure}
    \centering
    \includegraphics[width=\columnwidth]{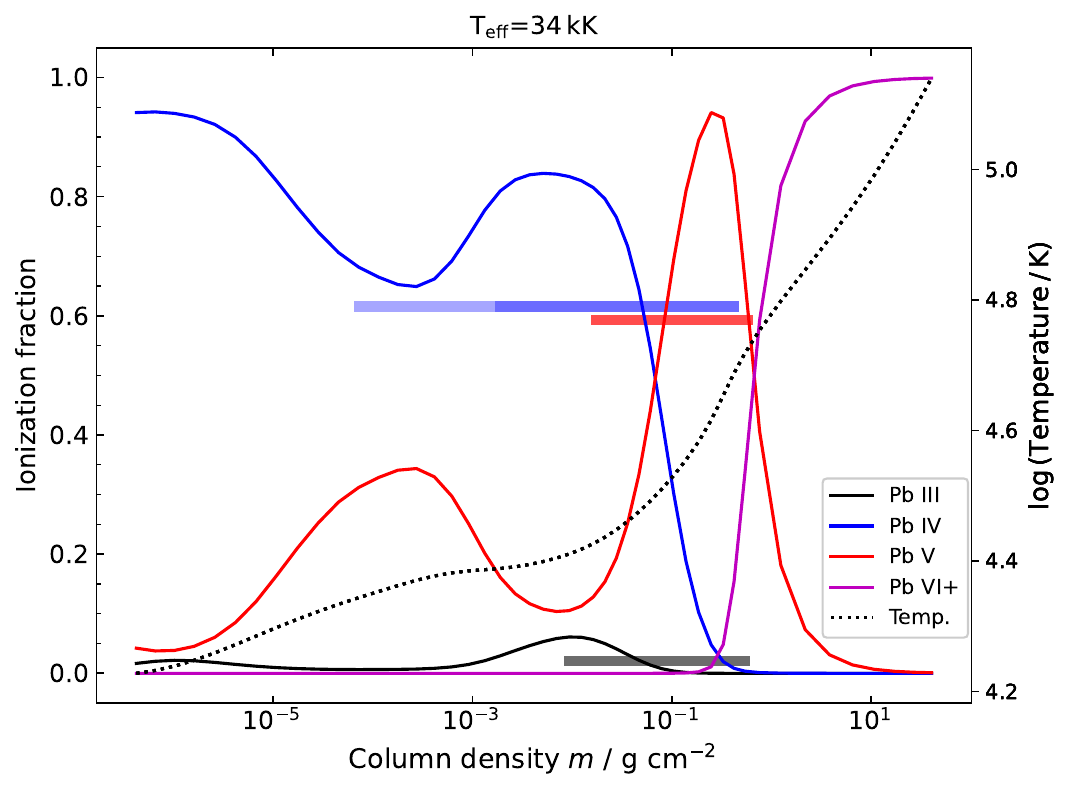}
    \includegraphics[width=\columnwidth]{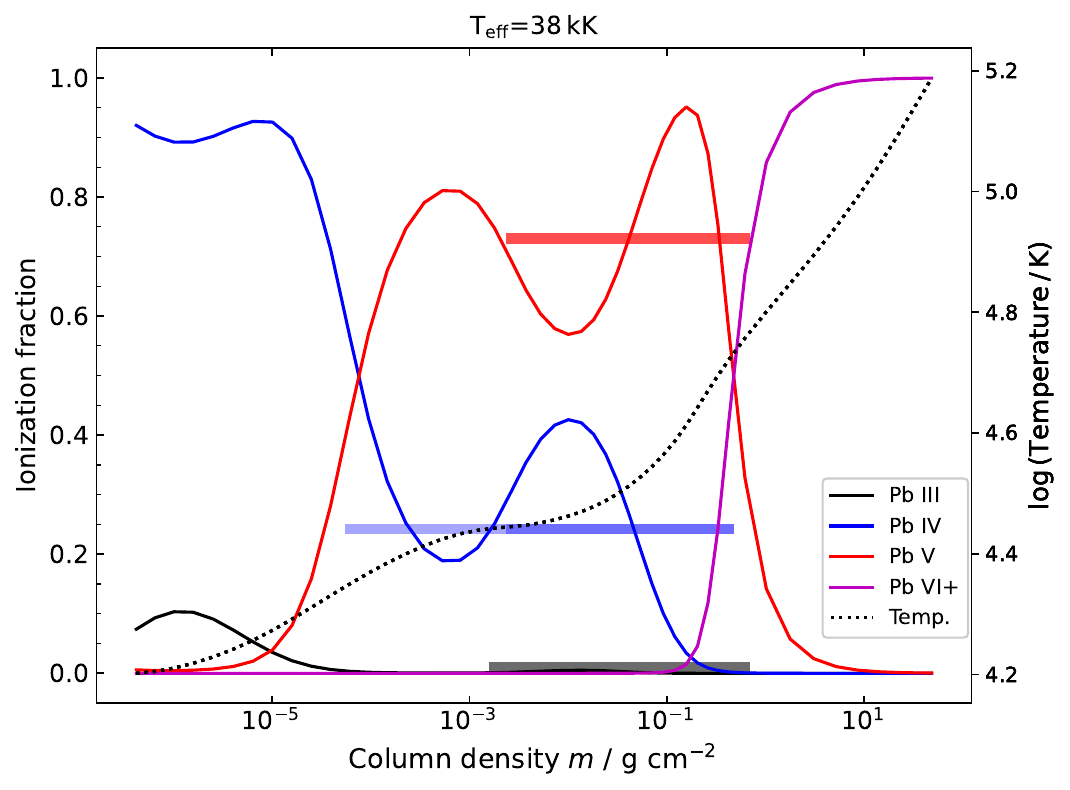}
    \includegraphics[width=\columnwidth]{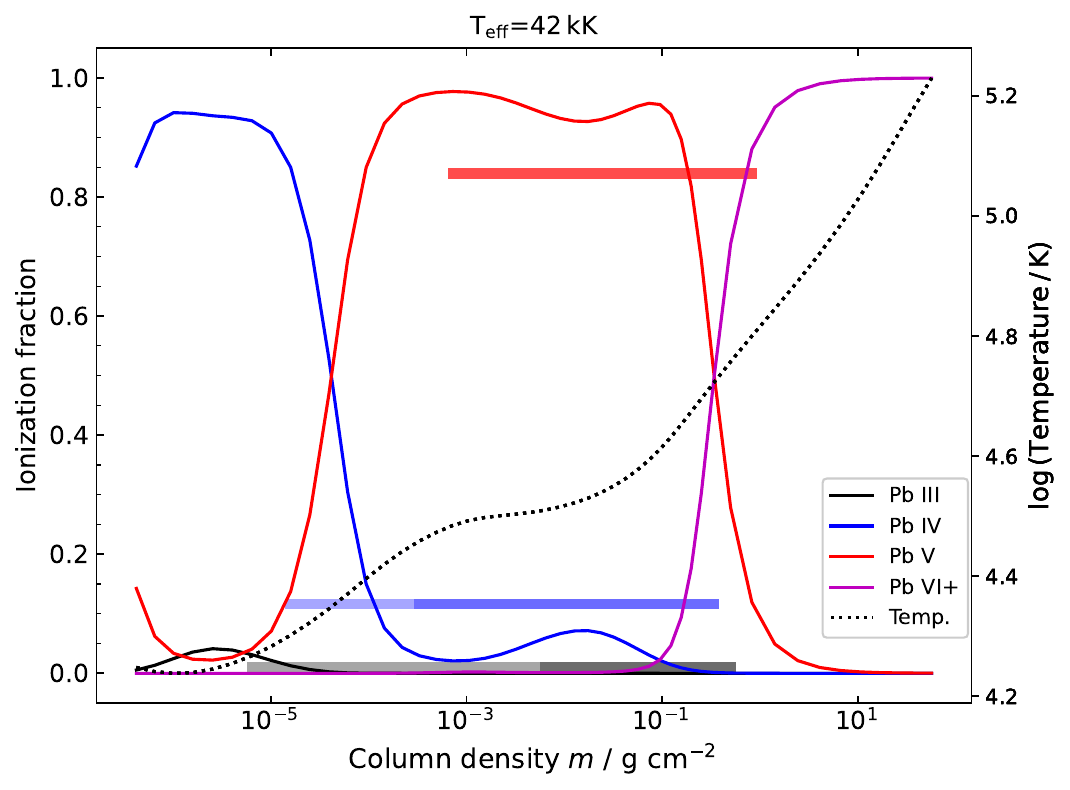}
    \caption{Ionisation fractions of lead for 34, 38 and 42\,kK models (top, middle and bottom panels), along with logarithm of temperature on the right-hand axis. The line labelled `Pb\,VI+' shows the fraction of all ionisation stages higher than Pb\,{\sc v}. The horizontal bars show the range in column mass in which lead should be present in order to produce lines with close to maximum equivalent width. These bars range between the shallowest value of $m_{\rm 0}$ and the deepest value of $m_{\rm 1}$ for the cusp positions of the strongest lines (maximum equivalent width higher than 2\,m\AA). The more opaque sections of the bars exclude the resonance lines at 1028 and 1313\,\AA~ (Pb\,{\sc iv}), and 1048\,\AA~ (Pb\,{\sc iii} at $T_{\mathrm{eff}}=42$\,kK).}
    \label{fig:ion}
\end{figure}

The change in line strength with temperature is a consequence of the ionisation fractions. These are shown according to the Saha equation in Fig.\,\ref{fig:ion}. In this figure, the region in which lines are formed for each ion is estimated by the horizontal bars. This bar spans the range between the minimum of \mzero{} and the maximum of \mone{} at the cusps of lines with maximum equivalent width greater than 2\,m\AA{}. The more opaque region of the bar excludes resonance lines from this range. Fig.\,\ref{fig:ion} shows that Pb\,{\sc v} is usually the dominant ion in the line-forming region, although Pb\,{\sc iv} is also abundant in the 34\,kK case. The resonance lines provide opacity higher in the atmosphere than the other lines, as shown by the more transparent extension to the horizontal bars in Fig.\,\ref{fig:ion}.

The approximate maximum extent of the line-forming region of resonance lines, shown by the more transparent horizontal bars in Fig.\,\ref{fig:ion}, can help to explain the change in cusp position of these lines with temperature, as seen in Fig.\,\ref{fig:spot}. At $T_{\rm{eff}}=34$\,kK, only the two Pb\,{\sc iv} resonance lines are separate from the main cloud of cusp positions. At $T_{\rm{eff}}=38$ and 42\,kK, they are joined by the two Pb\,{\sc iii} resonance lines.

\section{Discussion}
\label{sec:discussion}
In this study, we found that the response of each lead line to abundance stratification could be characterised by the location of the cusp, the narrowest enriched layer needed to produce the maximum equivalent width, in the \mzero-\mone{} plane. Per definition, any extension of the lead enrichment beyond the cusp has no further effect on equivalent width. The existence of the cusp can be understood in terms of the radiative transfer equation, in particular the role of extinction in different optical depth ranges.

To illustrate this, consider the equation for the monochromatic radiative flux, $F_{\nu}$, emitted from the surface of the star \citep[e.g.][]{gray75}:
\begin{equation}
    \label{eq:flux}
    F_{\nu} = 2\pi \int_0^{\infty} S_{\nu}(\tau_{\nu})E_2(\tau_{\nu}) d\tau_{\nu}.
\end{equation}
$S_{\nu}$ is the source function and $E_2$ is the extinction factor, both functions of optical depth, $\tau_{\nu}$. The subscript $\nu$ indicates a quantity taken at a specific frequency. 

The effect of lead opacity on the emitted flux is expressed through $E_2$, which is proportional to $\rm{e}^{-\tau_{\nu}}$. The presence of a lead layer causes the optical depth to increase more quickly with depth relative to $S_{\nu}$ than the lead-free case, hence reducing the flux. However, $S_{\nu}$ and $E_2$ have opposite behaviour with increasing $\tau_{\nu}$; $S_{\nu}$ increases sharply towards higher $\tau_{\nu}$, whereas $E_2$ asymptotes from unity at the stellar surface to zero at high $\tau_{\nu}$. This means that the atmosphere can be split into three regions:
\begin{enumerate}
    \item $\tau_{\nu}<<1$: Very little extinction ($E_2\approx1$). Eq.\,\ref{eq:flux} is dominated by $S_{\nu}$, which is unaffected by the presence of a lead layer.
    \item $\tau_{\nu}>>1$: High extinction ($E_2\approx0$). Whether  extra line opacity from lead is present or not, contributions to $F_{\nu}$ from this region are minimal. 
    \item $\tau_{\nu}\sim1$: Both $S_{\nu}$ and $E_2$ contribute to the flux integral. Here, the effect of extra line opacity on $E_2$ will significantly reduce the overall flux, causing an absorption line.
\end{enumerate}

The position of the cusp in the \mzero-\mone{} plane corresponds to the line-forming region in the atmosphere. The extension of the lead layer beyond the cusp does not strengthen the spectral line because the extension falls into regions (i) or (ii), where the effect of line opacity is minimal.

The exact extent of the line forming region, and hence the position of the cusp, is modulated by the weighted oscillator strength and the number of absorbers, which govern the ratio of line to continuum opacity. As these quantities increase, the line-forming region also becomes larger as the extinction at these wavelengths is greater.

The box-shaped lead profiles used in this study are a simplification of the stratification that could be produced by concentration of lead via radiative levitation. If a real star were to have a lead-enriched layer with a box-like profile in its atmosphere, the stratification could be represented by a single point in the \mzero-\mone{} plane. Fig\,\ref{fig:line_profiles} shows that the lead lines behave differently to an homogeneous atmosphere depending on the position of this point, with a deep layer creating absorption in the line wings, and a shallow layer absorbing in the core. We have investigated the effect of a Gaussian abundance distribution on the line profiles in App.\,\ref{app:gauss}, and find that the Gaussian and box profiles behave similarly when they are of similar width.

In addition, there may be differences in the shapes of lines in the optical and the UV if the point in the \mzero-\mone{} plane is near where the line cusps cluster. The general wavelength trend is that the cusp for longer wavelength lines occurs at lower \mone{} than for shorter wavelengths.
Therefore, if the optical lines in an observed spectrum appear normal but the UV lines are weaker than expected, this could be produced by stratification in the middle of the cloud of cusp positions in Fig.\,\ref{fig:spot}. Such a layer extends deep enough to encompass the cusps of most optical lines, but not most UV lines. In this case, the derived lead abundance would be higher in the optical compared to the UV, since the UV lines would only have lead in part of their line-forming regions. It should be noted that the cores of the strongest lines, such as the Pb\,{\sc iv} resonance line at 1313\,\AA{}, are affected by non-LTE effects; improved model atoms with photoionisation cross-sections are needed to fully model the line profiles.

For the 34\,kK model, many Pb\,{\sc v} lines also have deeper \mzero{} at the cusp than the optical lines; hypothetically, this could also allow a stratification profile near the cusps of Pb\,{\sc v} lines to be identified by normal Pb\,{\sc v} lines and broad, shallow optical lines, although no lead stars have been found at this low temperature. In any case, difficulty in finding a single abundance to fit all lines may be an indication of stratification, as found previously for Ap stars \citep{ryabchikova02}. Lastly, if all lines appear the same as an homogeneous abundance profile, then the atmosphere could be either not stratified at all, or the enriched layer could encompass the entire line-forming region.

The Pb\,{\sc iv} line at 4049\,\AA{} is potentially important for explaining the lack of detection of lead stars in the lower temperature range of heavy metal subdwarfs. This line is the most important line in the optical for lead stars, so an optical spectrum that does not show it precludes the star from the category. Fig.\,\ref{fig:spot}, on which the 4049\,\AA{} line is marked, shows the change in its cusp position with temperature; the cusp moves to  lower \mone, i.e. the bottom of the layer moves higher in the atmosphere, as temperature increases. If lead stars contain an enriched layer which is near the cusp at 42\,kK, i.e. $\log{m_{\mathrm 0}}\approx-3$, $\log{m_{\mathrm 1}}\approx-1$, then the 4049\,\AA{} line would increase in strength with temperature. However, stratification also likely varies with effective temperature. Thus more detailed atmosphere models including diffusion would be needed to explain the lack of cooler lead stars.

Figure\,\ref{fig:spot} indicates that in the region around the cluster of cusp positions, $\sim10^{-9}$\,M$_{\odot}$ of material is required to enrich the corresponding layer. This is equivalent to concentrating lead from a column mass of $\sim10^3$ to $10^4$\,g\,cm$^{-2}$ upwards. This is deeper than the lower boundary of our model atmospheres, which only contain around $10^{-11}$\,M$_{\odot}$ of total mass, assuming $R_*=0.13\,$R$_{\odot}$. The radial extent over which diffusion would need to act to incorporate the $10^{-9}$\,M$_{\odot}$ of material required for a 4 dex enhancement also depends on the structure below the atmosphere. More detailed models including diffusive transport are needed to answer this question.

Whilst we have only considered vertical abundance stratification in this paper, other chemically peculiar stars show horizontal variation in abundances over their surface \citep[e.g.][]{kochukhov04,alecian07,hummerich18}. As yet, no such evidence has been found for heavy metal stars. \citet{dorsch21} used time-resolved spectroscopy to measure abundances and radial velocities for EC 22536-5304, but found no change in elemental abundances with time (Dorsch, private communication).
Similarly, \citet{martin19.phd} analysed time series spectroscopy of LS\,IV--14$^\circ$\,116 over four half-nights, finding radial velocity variations attributed to g-mode pulsation. No time variation in equivalent width was found. A strong surface magnetic field  was ruled out for LS\,IV--14$^\circ$\,116 \citep{randall15}. It appears that any magnetic field may not be strong enough in the heavy-metal subdwarfs to influence an abundance stratification such as that seen in HgMn stars (e.g. Makaganiuk et al. 2012).

The cause behind the difference in effective temperature between the Pb-rich and Zr-rich stars is still an open question. We expect essentially similar results for the behaviour of Zr lines with vertical abundance stratification, i.e. Zr layers smaller than the line-forming region will be distinct in shape compared to the homogeneous case. However, all the known Zr stars pulsate \citep{ahmad05a,latour19a,ostensen20}; disentangling the effects of pulsation and stratification on the spectral line profiles adds another layer of complexity not present in most of the Pb stars. Ultimately, the temperature difference between the two groups will best be understood using atomic diffusion modelling, towards which work is ongoing.

\section{Conclusions}
\label{sec:conclusions}
In this study, we have used model atmospheres with parameters typical for a heavy metal subdwarf ($T_{\rm{eff}}$ between 34 and 42\,kK, $\log{g/\mathrm{cm\,s^{-2}}}=5.80$ and a hydrogen number fraction of 50 per cent) to study the effect of stratified lead abundance on the spectral line profiles. We have shown that position and size of the narrowest lead layer which produces maximum equivalent width, the cusp, varies for each line. In addition, the position of the cusp follows a wavelength trend - see the clustering of the cusps of longer wavelength lines towards lower \mone{} in Fig.\,\ref{fig:spot}.

We have shown that an atmosphere with lead concentrated in a narrow layer may be distinguishable observationally from an atmosphere with homogeneous lead abundance. This is due to lines formed from layers deeper in the atmosphere being subject to stronger broadening compared to those formed from higher lead layers. This contrasts an ordinary line formed in a chemically homogeneous atmosphere, which experiences both weak broadening in its core and stronger broadening in its wings. Thus, observed spectral lines which are broader or narrower than their equivalent width would suggest when compared to homogeneous model atmospheres could be an indication of chemical stratification. The most important lines for diagnosing stratification are the resonance lines, which have a large line-forming region. The Pb\,{\sc v} lines are also important, since they are formed deeper than the other lines, and thus are most sensitive to the position of the lower boundary of the enriched layer. All these lines are found at wavelengths less than 3000\,\AA{}.

An important marker for stratification is a difference in measured lead abundance between optical and UV lines. This is because the cores of resonance lines in the UV are formed in the outer parts of the photosphere, a region that does not affect the strength of optical transitions. Not only their strength, but also the shape of these resonance lines is sensitive to a large range of lead layer configurations. This combination of sensitivity in both strength and shape makes UV resonance lines the most important tool to study lead stratification with observed spectra.

Since the resonance lines are located in the UV, and the wavelength trend of other lines is only apparent across the entire UV to optical, observations across a broad wavelength range will be required to identify chemical stratification. UV observations of several heavy metal stars will address this question.

\section*{Acknowledgements}
The authors are indebted to the UK Science and Technology Facilities Council via UKRI Grant No. ST/V000438/1 for grant support. The Northern Ireland Department for Communities funds the Armagh Observatory and Planetarium (AOP).

\section*{Data Availability}
The models used in this investigation will be made available upon reasonable request to the authors.



\bibliographystyle{mnras}
\bibliography{strati} 




\appendix
\section{Gaussian abundance profiles}
\label{app:gauss}

\begin{figure}
    \centering
    \includegraphics[width=\columnwidth]{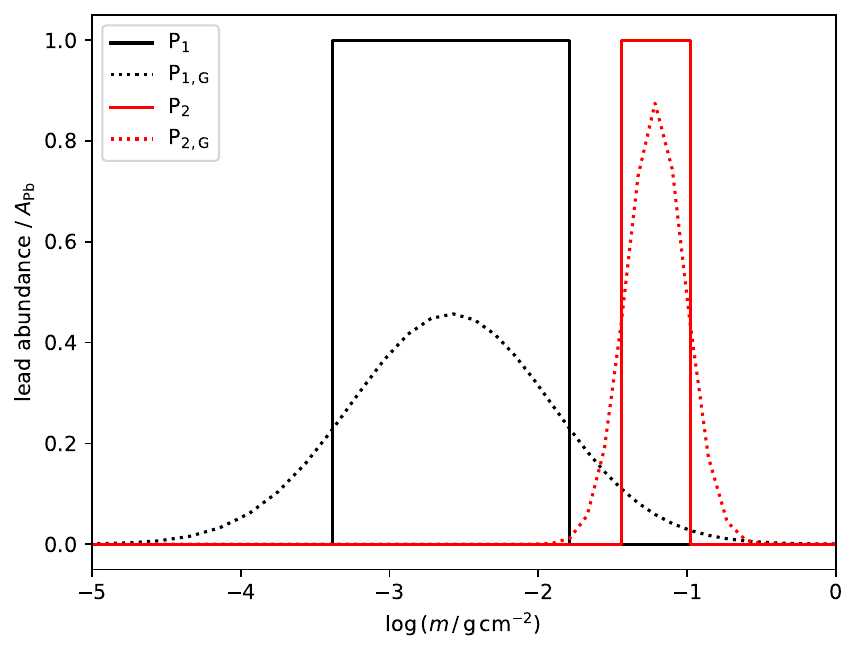}
    \caption{Lead abundance profiles P$_{\rm 1}$ and P$_{\rm 2}$ (as in Fig.\,\ref{fig:line_profiles}) and their corresponding Gaussian profiles, P$_{\rm {1,G}}$ and P$_{\rm {2,G}}$. The box profile lead abundance, $A_{\rm{Pb}}$, is $10^4$ times solar.}
    \label{fig:gauss}
\end{figure}

We have calculated some line profiles using a Gaussian vertical distribution of lead in order to test the effect of a smoother stratification profile. We chose Gaussian profiles in log column mass with a full width half maximum equal to the width of the corresponding box profile. The height of the Gaussian profile was chosen such that the enclosed mass was the same as its corresponding box profile. We have used the abundance profiles P$_{\rm 1}$ and P$_{\rm 2}$, as in Sec.\,\ref{sec:profiles}, as examples. These profiles are shown in Fig.\,\ref{fig:gauss}.

\begin{figure}
    \centering
    \includegraphics[width=0.48\textwidth]{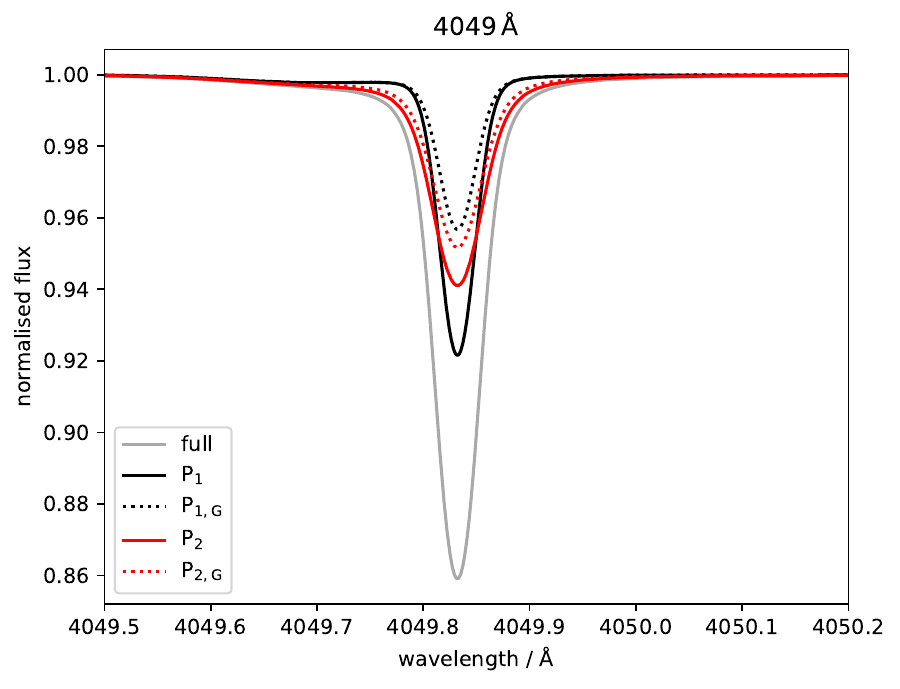}\\
    \includegraphics[width=0.48\textwidth]{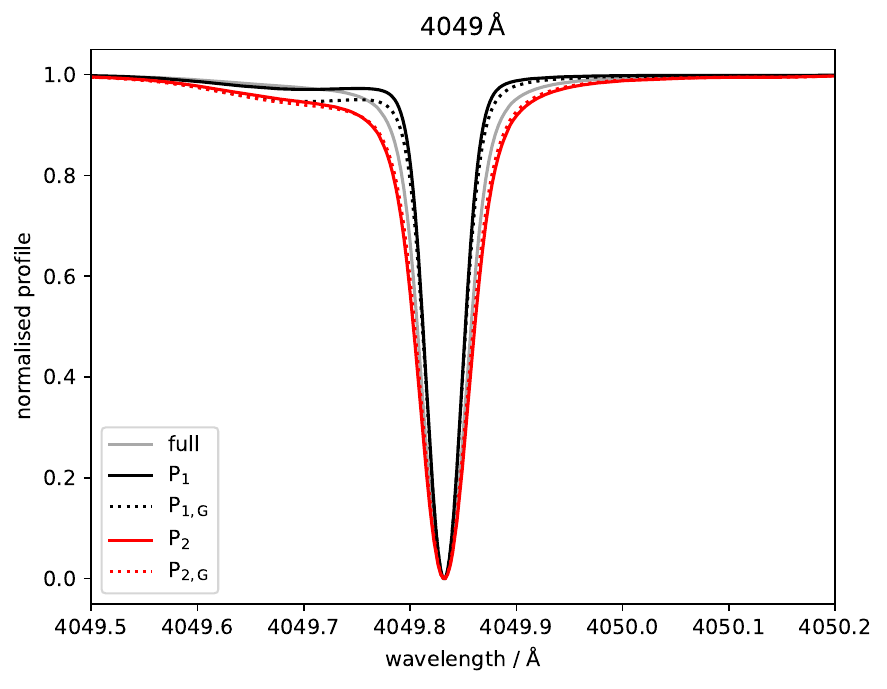}
    \caption{Comparison of spectral line shapes for 4049\,\AA{} produced using box abundance profiles (P$_{\rm 1}$ and P$_{\rm 2}$) and Gaussian abundance profiles (P$_{\rm {1,G}}$ and P$_{\rm {2,G}}$). The full spectral line profile, produced by an homogeneous lead abundance, is also plotted in grey. The top plot shows the absolute depth of the line below the continuum, whereas the line profiles in the bottom plot are normalised such that each line profile has the same depth.}
    \label{fig:gauss_4049}
\end{figure}

For most spectral lines, the line shape does not differ strongly between box and Gaussian abundance profiles. This is shown in  Fig.\,\ref{fig:gauss_4049}, in which the Pb\,{\sc iv} line 4049\,\AA{} is plotted for the different abundance profiles. The line depth varies, but this can be attributed to the Gaussian profiles having less lead mass in the centre of the line forming region compared to the box profiles. On the bottom panel of Fig.\,\ref{fig:gauss_4049}, the line depth is normalised; this allows the spectral line shapes to be compared directly. The line shape from the Gaussian profiles is just as distinct from the homogeneous abundance case as are the box profiles. This is particularly apparent from the red wing of the spectral line, since the blue wing contains a weak carbon feature unaffected by the lead abundance profile.

\begin{figure}
    \centering
    \includegraphics[width=\columnwidth]{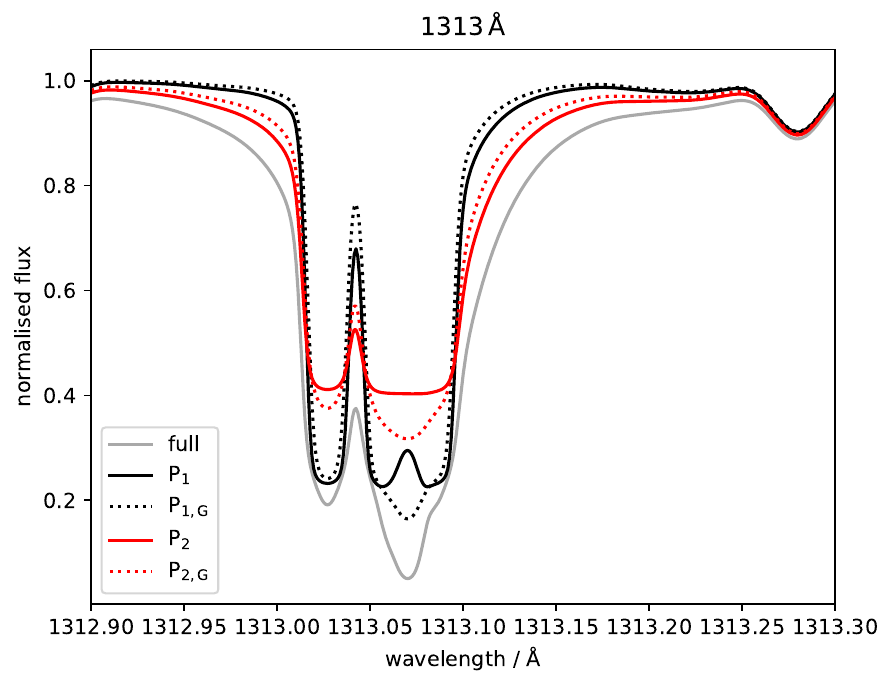}
    \caption{Comparison of spectral line shapes for 1313\,\AA{} produced using box abundance profiles (P$_{\rm 1}$ and P$_{\rm 2}$) and Gaussian abundance profiles (P$_{\rm {1,G}}$ and P$_{\rm {2,G}}$). The full spectral line profile, produced by a homogeneous lead abundance, is also plotted in grey.}
    \label{fig:gauss_1313}
\end{figure}

Fig.\,\ref{fig:gauss_1313} shows spectral line shapes for 1313\,\AA{}. This line is a blend of components (see Table\,\ref{tab:lines}); the strongest of these at 1313.07 behaves differently in the box and Gaussian profiles, due to saturation effects in the box profile case (see discussion of Fig.\,\ref{fig:line_profiles} in Sec.\,\ref{sec:profiles}). The lead is less concentrated in the Gaussian profile, so it behaves more similarly to the homogeneous case.


\bsp	
\label{lastpage}
\end{document}